%% file: sigconf.tex
\documentclass{acmart}
\usepackage{tabularx}
\usepackage{booktabs}
\usepackage{array}
\usepackage{xcolor}
\definecolor{modelGold}{HTML}{805B00}
\definecolor{modelBlue}{HTML}{175A96}
\definecolor{modelOrange}{HTML}{A84408}
\definecolor{modelPurple}{HTML}{713A99}
\definecolor{modelRed}{HTML}{B33025}
\definecolor{modelGray}{HTML}{555555}
\AtBeginDocument{%
  }

\setcopyright{acmlicensed}
\copyrightyear{2018}
\acmYear{2018}
\acmDOI{XXXXXXX.XXXXXXX}
\acmConference[Conference acronym 'XX]{Make sure to enter the correct
  conference title from your rights confirmation email}{June 03--05,
  2018}{Woodstock, NY}
\acmISBN{978-1-4503-XXXX-X/2018/06}

\begin{document}

\title{From Instructional Intent to Classroom-Ready Lessons: Investigating Teacher–AI Collaboration in Lesson Authoring for Secondary Computing Education}

\author{Tahreem Yasir}
\correspondingauthor
\email{tyasir@ncsu.edu}
\orcid{0001-6505-1915}
\author{Wenbo Li}
\email{wli55@ncsu.edu}
\author{Deniz Ozturk}
\email{dozturk@ncsu.edu}
\author{Yasitha Rajapaksha}
\email{yrajapa@ncsu.edu}
\author{Dmitri Droujkov}
\email{ddroujk@ncsu.edu}
\author{Xiaoyi Tian}
\email{}
\author{Tiffany Barnes}
\email{tmbarnes@ncsu.edu}
\affiliation{%
  \institution{North Carolina State University}
  \city{Raleigh}
  \state{North Carolina}
  \country{USA}
}
\renewcommand{\shortauthors}{Yasir et al.}


\begin{abstract}
Generative AI is increasingly studied as a tool for supporting lesson authoring; the creation and adaptation of classroom materials, yet little is known about how it should fit into secondary computing teachers’ subject-specific practices. We conducted two complementary qualitative studies with eight secondary computing teachers: focus groups examining existing authoring practices and a think-aloud study using LessonBuilder, a structured AI-supported lesson authoring system, as a research probe. We found that teachers’ instructional intentions developed through creating, inspecting, and adapting materials, while teachers selectively delegated tasks to AI and retained responsibility for judging classroom readiness. The think-aloud sessions further showed how examining AI-generated materials prompted teachers to clarify and refine their intentions, making authoring an ongoing negotiation between what teachers intended and what the AI produced. In the context of computing lessons, classroom readiness also involved considering how programming materials and supporting guidance matched students’ understanding and intended learning goals. We synthesize these findings into a conceptual model explaining how teachers coordinate instructional intent, AI assistance, and judgments of classroom readiness during lesson authoring. We derive design recommendations for supporting this coordination while preserving teacher control and addressing computing-specific authoring needs.

\end{abstract}


\keywords{Do, Not, Use, This, Code, Put, the, Correct, Terms, for,
  Your, Paper}

\received{20 February 2007}
\received[revised]{12 March 2009}
\received[accepted]{5 June 2009}

\maketitle
\input{01_Intro}

\input{02_rw}
\input{03_system}

\input{04_methods}
\input{05_results_S1}

\input{05_results_S2}
\input{06_framework}

\input{07_discussion}

\input{08_conclusion}

\begin{acks}
To Robert, for the bagels and explaining CMYK and color spaces.
\end{acks}

\section*{Ethics and Privacy Statement}
The research received institutional review board (IRB)
approval. Participation was voluntary, and teachers could
decline or withdraw from the research. We took measures to
protect participant confidentiality throughout data handling
and reporting. Teachers were represented using participant
identifiers, and identifying details were removed from
reported excerpts. Study records, including recordings,
transcripts, interaction logs, and authored materials, were
handled confidentially, with access restricted to the
research team.

\bibliographystyle{ACM-Reference-Format}
\bibliography{sigconf,sample-base}

\appendix

\end{document}

%% file: 01_Intro.tex
\section{Introduction}
\label{sec:introduction}

Teachers across subject areas translate curricular requirements into instructional material to fit their classroom context and learner need \cite{kaufman_how_2020}. They draw on curriculum guides, instructional platforms, online resources, and previous lessons when authoring this materials \cite{rhodes_design_2018}. These resources provide useful starting points, but they may not match a particular class's learning goals, prior knowledge, pacing, or classroom conditions \cite{polikoff_challenges_2018,wang_teachers_2021}. To address these differences, teachers create, adapt, and combine explanations, examples, activities, assignments, and assessments \cite{sherin_curriculum_2009}.We call the process of developing a coherent set of materials aligned with a particular classroom's learning goals, students' needs, and teaching conditions \emph{lesson authoring}.

Generative AI is increasingly being explored as a means of supporting teachers' lesson authoring process \cite{cheah_integrating_2025,albaloul_systematic_2026}. Understanding this support involves considering both what AI generates and how teachers can direct and revise its contributions during authoring. General-purpose conversational systems support ongoing dialogue, but do not typically separate instructional planning from the development of lesson materials \cite{choi_exploring_2026,lehtinen_generative_2026,subramonyam_bridging_2024}. Recent work has explored more structured interactive authoring systems in which teachers can create and refine lesson materials with AI support \cite{fan_lessonplanner_2024,cui_codesigning_2026,kaputa_simstep_2026}.

The design of AI support for lesson authoring must also account for the demands of the subject being taught. In secondary computing education, teachers need to develop programming materials suited to students' prior knowledge \cite{reichert_framework_2027}, while anticipating the errors and debugging difficulties students may encounter when working with those materials \cite{vinnervik_-depth_2023}. Understanding how teachers make these decisions provides a basis for examining where AI assistance may fit into their authoring practices. However, existing lesson-authoring systems have largely been developed in broader or non-computing educational contexts, offering limited insight into these decisions in secondary computing classrooms \cite{fan_lessonplanner_2024,cui_codesigning_2026,calo_cat-ai_2026}. Although recent research has begun to explore generative AI for preparing programming- and AI-related instructional materials \cite{yu_improving_2026}, it remains unclear how secondary computing teachers use and direct AI assistance when creating, adapting, and judging materials for their students, and how these practices should inform the design of AI support.

To examine how AI fits within teachers' lesson-authoring work, we conducted two complementary qualitative studies with eight secondary computing teachers, bringing together accounts of existing practices and observations of authoring with AI support. Study~1 used focus groups to examine how teachers create and adapt lesson materials, use AI, and judge whether materials suit their classroom's learning goals, students, and teaching conditions. Study~2 examined teachers' authoring decisions through individual think-aloud sessions using LessonBuilder as a research probe. We developed this interactive lesson-authoring system based on prior research \cite{fan_lessonplanner_2024,kaputa_simstep_2026,cui_codesigning_2026}. LessonBuilder allows teachers to inspect and revise AI-generated plans before generating lesson materials and to request selective adaptations. As teachers worked with the system, we examined how they expressed and refined their instructional intentions, directed AI assistance, and evaluated the resulting materials. Together, the studies allowed us to consider teachers' accounts of their everyday authoring work alongside the decisions they made when structured AI support was available. They addressed the following research questions:

\textbf{RQ1:} How do secondary computing teachers create and adapt lesson materials, judge their suitability for classroom use, and incorporate AI into their existing authoring practices?

\textbf{RQ2:} How do secondary computing teachers articulate and refine instructional intent and exercise control during lesson authoring with the structured AI support provided by LessonBuilder?

\textbf{RQ3:} How can findings across both studies inform a conceptual model of teacher-orchestrated AI-supported lesson authoring and corresponding design recommendations?

Our central contribution is a conceptual model of \emph{teacher-orchestrated AI-supported lesson authoring}, grounded in findings across both studies with secondary computing teachers. The model describes how teachers decide what to delegate to AI and where to retain control as they refine instructional intent, create and adapt materials, and judge classroom readiness. We situate this model in relation to existing frameworks of curriculum use and AI-supported assessment authoring \cite{sherin_curriculum_2009,cui_codesigning_2026}. Drawing on the findings and model, we offer four design recommendations: support selective revision, keep evolving lesson requirements explicit and revisable, support teachers in evaluating AI feedback, and enable inspection of both the technical and instructional aspects of computing materials.

%% file: 02_rw.tex
\section{Related Work}
\subsection{Teachers’ Use and Adaptation of Curriculum Materials}


Teachers use curriculum materials, including textbooks, worksheets, lesson plans, and digital resources, to translate curricular goals into instruction \cite{kaufman_how_2020}. They obtain materials from schools, districts, online sources, and colleagues, or develop their own \cite{rhodes_design_2018}. However, available materials may require supplementation or adaptation to address the required topics and learning objectives and suit students' needs \cite{kaufman_how_2020,choppin_role_2022,polikoff_challenges_2018}. Teachers assess whether these materials meet curricular and institutional expectations, engage students, offer an appropriate level of challenge, and are practical to use in their classrooms \cite{wang_teachers_2021,polikoff_identifying_2021}. These assessments guide changes to content and presentation, including combining materials from different sources to fit students and instructional goals \cite{kaufman_how_2020,sherin_curriculum_2009,reichert_framework_2027}. Teachers also evaluate the resulting materials before classroom use, considering whether the revised content meets their instructional needs \cite{polikoff_supplemental_2019}.

In K-12 computing education, teachers must consider whether materials explain computing concepts clearly and provide programming activities suited to their students. Materials may also require revision as the tools used in instruction and the technologies students are expected to learn change \cite{reichert_framework_2027}. When authoring programming activities, teachers must consider the concepts students need to understand beforehand, whether the code works as intended, and the errors and debugging difficulties students may encounter \cite{vinnervik_-depth_2023,gale_hands-up_2026}. Differences in students' prior programming experience may require teachers to adapt activities and provide different levels of guidance \cite{reichert_framework_2027,gale_hands-up_2026}.
These decisions draw on teachers' knowledge of programming and how to teach it, making their preparation and access to support relevant to authoring. Research on K--12 computing education has identified challenges in teacher availability and preparation \cite{wilson_running_2010,zhang_introducing_2025}. Studies also report variation in teachers' computing expertise \cite{czemerinski_whats_2020,kim_landscape_2021} and limited access to peer support \cite{ni_computer_2023}. These findings motivate support that helps computing teachers develop and adapt materials while evaluating their suitability for students and classroom use.

\subsection{Generative AI for Teacher Work and Lesson Authoring}

Generative AI has been explored to provide support within this work of authoring and adapting lesson materials. Researchers have explored how AI can assist teachers in developing exercises \cite{calo_cat-ai_2026} and designing learning activities \cite{wang_enabling_2025}. This support also includes opportunities to shape content during generation: SimStep allows teachers to inspect and revise representations of concepts, scenarios, and learning goals when authoring interactive STEM simulations \cite{kaputa_simstep_2026}. LessonPlanner connects goal refinement, outline generation, and section-level development and revision, drawing on pedagogical guidance to support lesson planning \cite{fan_lessonplanner_2024}. Other work addresses learner differences: AdaPT supports transforming lesson plans using student profiles \cite{zhang_adapt_2026}, FACET uses a multi-agent architecture to personalize worksheets \cite{gonnermann-muller_facet_2026}. Shiksha Copilot and Ripplet provide starting points for customization and reuse of materials \cite{varuvel_dennison_shiksha_2026,cui_codesigning_2026}. Ripplet additionally conceptualizes assessment authoring as an iterative process in which teachers develop assessments while refining their requirements \cite{cui_codesigning_2026}. These approaches support both material production and teachers' ongoing involvement in shaping instructional content.

However, materials generated by AI may still require teacher review before classroom use. Generated content may fit poorly within a lesson's context \cite{yu_improving_2026}, and the instructional content and activities may require further revision for pedagogical suitability \cite{kerr_prompts_2025,yu_improving_2026}. Teachers must assess both the information in generated materials and whether those materials meet their instructional goals and students' needs
\cite{cui_codesigning_2026,kaputa_simstep_2026}. Teacher review involves checking and correcting information, comparing alternatives, and revising selected content before deciding what to incorporate into the lesson \cite{yu_improving_2026,cui_codesigning_2026,kalenda_pre-service_2025,kaputa_simstep_2026}. AI support thus leaves teachers with continuing responsibility for judging both material quality and classroom fit.

Building on this work, we examine how secondary computing teachers connect decisions about instructional intent, AI assistance, and classroom readiness during lesson authoring. We bring together their accounts of existing practices and observations of authoring with LessonBuilder to investigate how they refine their intentions, direct AI assistance, and judge the resulting materials.


\subsection{Teacher–AI Collaboration, Agency, and Control}

Understanding these authoring decisions requires considering how interaction with AI supports teacher control. Human–AI interaction research offers relevant perspectives through agency, transparency, and human authority.

\textbf{Agency} concerns users' ability to determine how work proceeds and steer AI assistance toward their objectives \cite{sharma_investigating_2024,holter_deconstructing_2024,lam_just--time_2026}. In lesson authoring, prior work highlights teachers' preferences for directing how AI contributes throughout the process \cite{lawrence_how_2024}. Systems support this involvement through opportunities to intervene at checkpoints and refine both instructional objectives and developing materials \cite{kaputa_simstep_2026,varuvel_dennison_shiksha_2026,fan_lessonplanner_2024,cui_codesigning_2026}.

\textbf{Transparency} supports informed judgment by making relevant information about system behavior and outputs accessible to users \cite{amershi_guidelines_2019,pareek_sensemaking_2026}. One way authoring systems support transparency is by exposing intermediate outputs for inspection before users proceed \cite{wu_ai_2022,pareek_sensemaking_2026,kaputa_simstep_2026}. For teachers, inspectable lesson outlines or representations of concepts and learning goals provide opportunities to judge whether the developing materials reflect their intentions and identify where revision is needed \cite{fan_lessonplanner_2024,kaputa_simstep_2026}.

\textbf{Human authority} concerns who has the final say over outputs produced with AI assistance \cite{chiang_are_2023,li_modeling_2023}. Positioning AI contributions as proposals for consideration preserves opportunities for users to decide whether to incorporate them \cite{yizhou_tian_understanding_2026,ma_towards_2025}. For teachers, this means deciding whether to retain, revise, or reject generated materials before they become part of a lesson \cite{lawrence_how_2024,cui_codesigning_2026}.

Together, these perspectives connect how teachers direct AI assistance, what they can inspect, and which decisions remain theirs. Examining these relationships during authoring can help explain how teachers communicate instructional intentions, respond when AI contributions need revision, and decide what to retain.




%% file: 03_system.tex
\section{LessonBuilder: A Teacher-Directed AI System for Multi-Phase Lesson Authoring}
\label{sec:system}

LessonBuilder is a teacher-directed, AI-supported system for creating and adapting computing lessons. It organizes authoring around intermediate artifacts and teacher-directed transitions rather than generating a complete lesson in one step. Teachers begin by describing their lesson goals, classroom context, and desired materials, or by importing an existing lesson for adaptation. The system presents the resulting outputs for review, and teachers choose whether to request revisions or continue with the current version.

\subsection{Design Rationale}
\label{sec:dr}

Three goals informed LessonBuilder's design. First, teachers should be able to create new lessons or adapt existing resources. Research describes curriculum preparation as selecting, adapting, and creating materials \cite{grave_teachers_2015,sherin_curriculum_2009}, motivating LessonBuilder's generation and import pathways.

Second, teachers should be able to inspect and revise materials as they develop. Following prior work on intermediate representations \cite{varuvel_dennison_shiksha_2026,fan_lessonplanner_2024,kaputa_simstep_2026}, LessonBuilder presents proposed lesson plans, import previews, generated materials, and quality review feedback as separate outputs during authoring. This design is intended to help teachers examine how explanations, programming examples, practice, and assessments fit together, and inspect code and test cases where available.

Third, teachers should retain control over consequential authoring decisions. Building on teacher-directed assessment creation and revision in Ripplet \cite{cui_codesigning_2026}, LessonBuilder lets teachers decide whether to proceed from a proposed plan to material generation, whether to apply quality review feedback, and when to publish the lesson. Teachers explicitly initiate these operations.

\subsection{Role-Specialized LLM Architecture}
\label{sec:architecture}

LessonBuilder uses separately invoked LLM agents for distinct authoring tasks. The generation pathway comprises a \emph{Planner}, a \emph{Generator}, and a \emph{Quality Reviewer}. Each agent is implemented as an LLM API call configured with a role-specific prompt and task-relevant context to return structured output. A deterministic orchestration layer invokes the appropriate agent in response to teacher actions and manages the inputs and outputs of each operation. The agents neither select subsequent operations nor communicate through an open-ended conversation. 

\begin{figure}
    \centering
    \includegraphics[width=\linewidth]{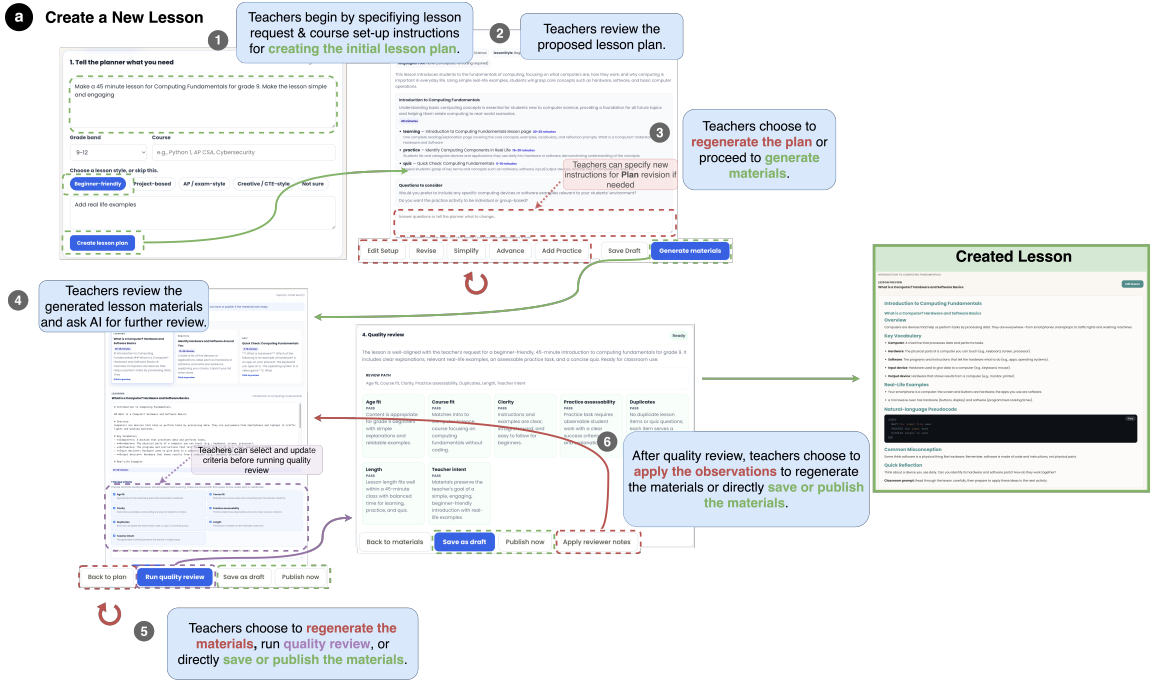}
    \caption{LessonBuilder's generation pathway. Teachers inspect and revise plans and materials (steps 1-3) before saving or publishing, with optional AI quality review (steps 4-6) to inform further revisions.}
    \label{fig:ui_1}
\end{figure}

Both pathways produce lessons organized into ordered topics containing learning, practice, and quiz items. Learning items contain explanations and programming examples (Figure~\ref{fig:ui_1}, Figure~\ref{fig:lessonbuilder-workflow2}). Practice items can include assignment instructions, starter code, hints, a model solution, and test cases, while quiz items contain questions, optional response choices, and expected answers. Teachers can also author these items manually (Figure~\ref{fig:lessonbuild_chatbot}(a)).

\subsubsection{Generation Pathway: Role-Specialized Agents}
The generation pathway (Figure~\ref{fig:ui_1}, step 1)  begins with the \emph{Planner}, which uses the teacher's lesson goals, classroom context, and desired materials to produce a structured plan. The plan specifies a lesson summary, topics, activities, assessments, and estimated durations, and may include questions for clarification. 

Once the teacher chooses to proceed, the \emph{Generator} (Figure~\ref{fig:ui_1}, step 3) uses the plan, course context, and any revision instructions to produce the learning, practice, and quiz items described above (Figure~\ref{fig}, steps~3-5). Subsequent revision requests invoke the same agent with updated instructions.

After generation, teachers can optionally invoke the \emph{Quality Reviewer}, (Figure~\ref{fig:ui_1} step 4-6) . Its input includes the current plan, generated items, and teacher-selected criteria; its output contains criterion-specific observations, an overall status, and suggested revisions. This feedback remains separate from the materials. If the teacher chooses to apply the review suggestions, LessonBuilder combines them into a single revision instruction and invokes the Generator again. This LLM-generated review provides advisory feedback rather than independent verification of correctness, pedagogical quality, or classroom readiness.

A persistent \texttt{LessonPlanningSession} connects these operations by storing the teacher's inputs, course context, current plan, materials, revision instructions, and review findings. For each teacher-initiated operation, a deterministic orchestration layer retrieves the required session data, constructs the selected agent's input, and stores its output. It also checks dependencies: material generation requires a plan, and quality review requires generated materials. Agents neither update the session directly nor select the next operation.

\subsubsection{Import Pathway: Importing Items from Existing Materials}
\label{sec:import-architecture}

The import pathway combines deterministic document processing with separate LLM operations for structure identification and item generation (Fig~\ref{fig:lessonbuilder-workflow2}). It accepts PDF and DOCX files, accessible Google documents, or pasted content. After text extraction, an LLM identifies the primary programming language, topics, and proposed learning, practice, and quiz items, specifying each item's type, title, and summary.

The system divides the extracted text into passages and uses embedding-based retrieval to select the four most relevant passages for each proposed item. An item-specific LLM call then uses these passages as source context to generate the item's content. Deterministic parsing and normalization assemble the topics and generated items into an editable import preview.

Item generation occurs before teachers inspect the preview and confirm the import. Teachers can revise the proposed topics and item content before confirmation. The preview is maintained separately from the \texttt{LessonPlanningSession} used in the generation pathway. Confirmed imports use the same learning, practice, and quiz item structure as generated and manually authored materials; publication remains a separate teacher-initiated action.

The current import pipeline does not generate test cases for practice items. Teachers can add these subsequently through the lesson editor.

\subsection{Teacher-Facing Authoring Workflows}
\label{sec:authoring-workflows}

\begin{figure}
    \centering
    \includegraphics[width=\linewidth]{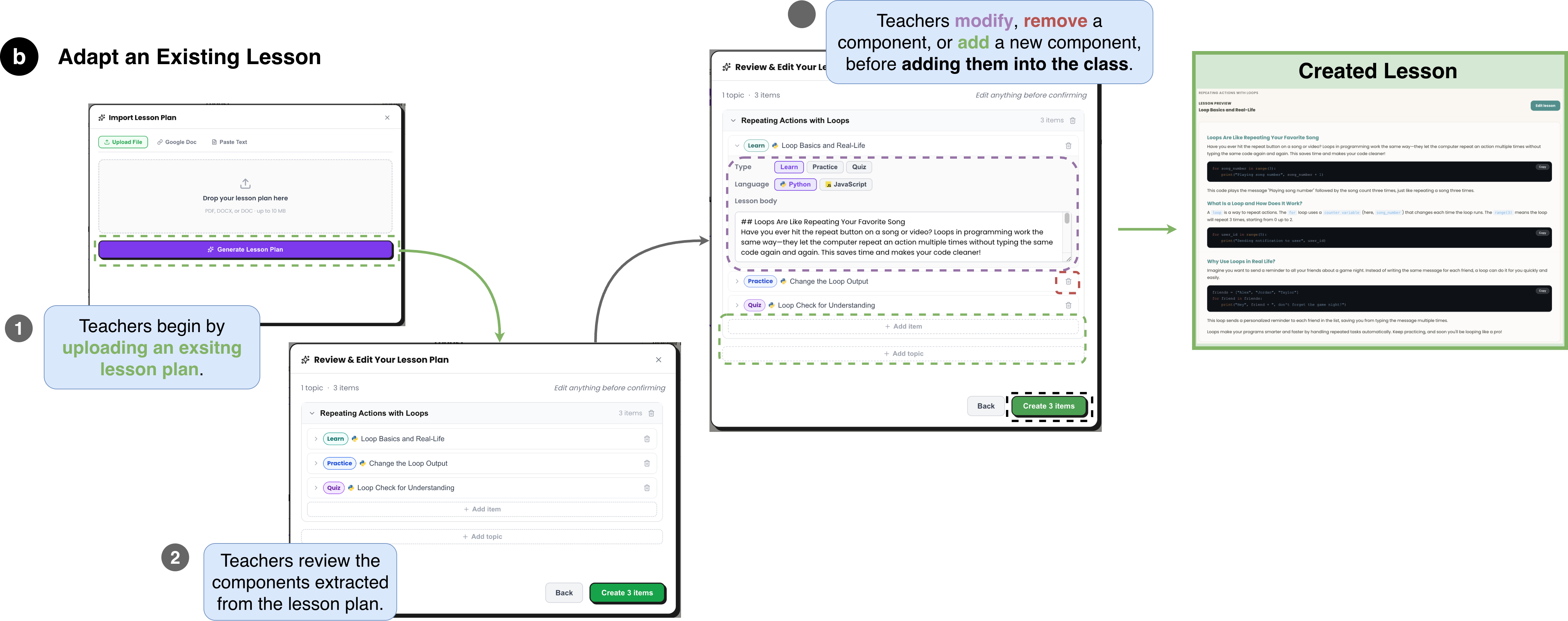}
    \caption{LessonBuilder's import pathway. Teachers upload existing materials (step 1), inspect the proposed topics and lesson items (step 2), and edit, add, or remove content (step 3) before confirming the import.}
    \label{fig:lessonbuilder-workflow2}
\end{figure}

Building on these architectural components, the interface lets teachers inspect and revise lesson content as they move through the generation and import pathways. Figure~\ref{fig:ui_1} and Figure~\ref{fig:lessonbuilder-workflow2} present the two AI-supported authoring pathways and their main decision points.

\subsubsection{Generating a New Lesson}

Teachers begin by describing the lesson's instructional goals, classroom context, and desired materials. The interface presents a proposed plan where teachers can inspect and revise topics, activities, assessments, and durations, before choosing to generate the materials (Figure~\ref{fig:ui_1}, step 1).

The generated lesson develops this plan into ordered topics containing learning, practice, and quiz items. Teachers can inspect explanations and programming examples alongside related activities and assessments. For programming assignments, this includes inspecting the proposed task and supporting code before deciding whether further revision is needed (Figure~\ref{fig:ui_1}, step 2-4).

Teachers may then request a Quality Review using selected criteria. The interface presents the findings separately from the lesson, allowing teachers to continue without applying them or request a revision. In the current system, this revision applies the available suggestions collectively rather than individually (Figure~\ref{fig:ui_1}, step 4-6).

\subsubsection{Importing and Adapting Existing Materials}

Teachers who already have lesson materials can begin with PDF or DOCX files, accessible Google documents, or pasted content. The interface presents an editable preview of the inferred topics and learning, practice, and quiz items. Teachers can inspect how the source has been organized and revise the proposed topics and item content before confirming the import (Figure~\ref{fig:lessonbuilder-workflow2}, step 1-2). 

For materials containing programming content, the preview can include explanations, programming examples, practice instructions, and starter code. Teachers can inspect how these elements have been represented within the proposed lesson items. The current import pipeline does not generate test cases for practice items; teachers can add these later in the lesson editor Figure~\ref{fig:lessonbuilder-workflow2}, step 3-4.

\subsubsection{Teacher Control Over Lesson Revision}

\begin{figure}
    \centering
    \includegraphics[width=\linewidth]{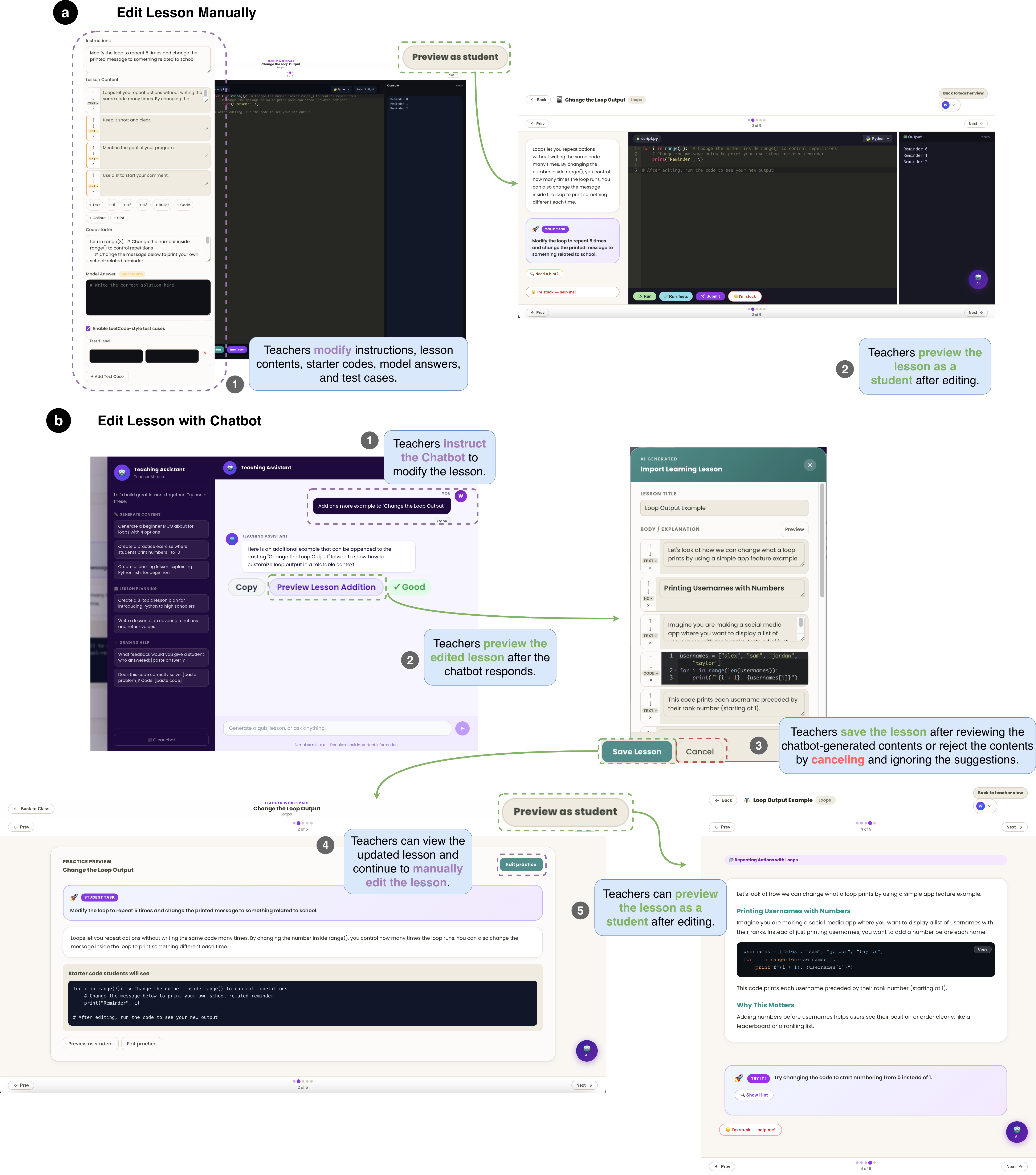}
    \caption{LessonBuilder's editing support - setting up programming activity (a) and lesson editing with chatbot (b)}
    \label{fig:lessonbuild_chatbot}
\end{figure}

After generating a lesson or confirming an import, teachers can save materials as a draft or publish them to preview their teacher- and student-facing presentations. Once the lesson is published, the lesson remains editable through manual controls and teacher-facing chatbot assistance (Figure~\ref{fig:lessonbuild_chatbot}(b)).

Teachers can manually edit individual items, reorder or remove them, and add materials directly. For programming activities, the editor exposes assignment instructions, starter code, hints, model solutions, and optional test cases. Teachers can revise these components or author them themselves, including adding tests to imported practice items.

The teacher-facing chatbot becomes available after the lesson is made available to students. It supports natural-language revision requests, such as ``Add a multiple-choice question about for loops with four options.'' The chatbot proposes lesson updates that teachers can approve or discard.

\subsection{Implementation Details}
\label{sec:implementation}

LessonBuilder uses a client--server web architecture. A React frontend, built with Vite, provides the authoring interface, editable artifacts, previews, and programming environment. A Node.js and Express backend handles authoring requests, coordinates LLM operations, processes imported documents, and manages saving and student access to lessons. MongoDB, accessed through Mongoose, stores accounts, classes, lesson materials, and authoring state. The system uses JSON Web Tokens for authentication.

The backend accesses OpenAI models through the OpenAI Node SDK. The Planner, Generator, and Quality Reviewer described above are role-specialized LLM calls that respectively propose lesson plans, generate lesson materials, and provide advisory feedback using teacher-selected criteria. The Planner, Generator, Quality Reviewer, import-structuring operation, imported-item generator, and teacher-facing chatbot use \texttt{gpt-5.2}, with operation-specific prompts, input contexts, response structures, and generation parameters. The Planner uses a temperature of 0.2 and a 3,000-token limit; the Generator uses 0.4 and 6,000 tokens; and the Quality Reviewer uses 0.1 and 2,500 tokens. Import structuring uses a temperature of 0.1 and a 6,000-token limit, while imported-item generation uses 0.3 and 10,000 tokens.

For document import, \texttt{pdf-parse} extracts PDF text and Mammoth processes DOCX files. Embedding-based retrieval uses OpenAI's \texttt{text-embedding-3-small} to identify source passages relevant to each proposed item.

The programming environment uses Ace Editor and supports Python and JavaScript. Python programs execute in the browser through Skulpt, while JavaScript programs execute through the browser's JavaScript \texttt{Function} constructor. The interface captures program output and runtime errors.

%% file: 04_methods.tex
\section{Methods}
\label{sec:methods}

\begin{figure*}[t]
\centering
\includegraphics[width=\textwidth]
{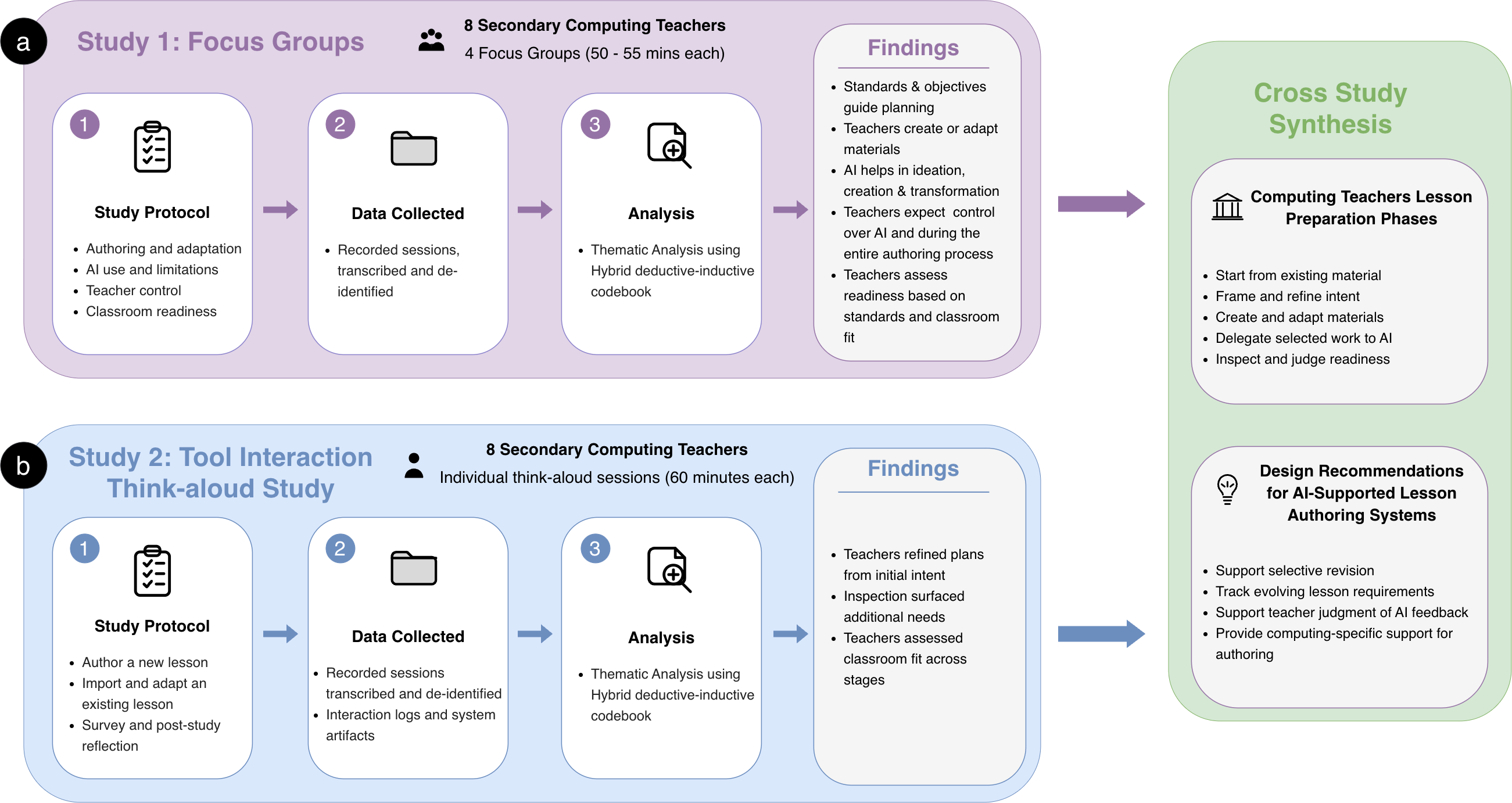}
\caption{Study Design}
\label{fig:study-design}
\end{figure*}

We conducted two complementary IRB-approved qualitative studies during the final week of a four-week professional development (PD) workshop for secondary computing teachers at a public university in the southeastern United States.
 Study~1 explored teachers' lesson-authoring and
adaptation practices, their use of AI, and their expectations for
control over AI. Study~2 examined how the same teachers expressed
instructional intent and exercised control while completing
authoring tasks using LessonBuilder's staged workflows. Study~1 did not inform the design of LessonBuilder or Study~2. Figure~\ref{fig:study-design} describes the two study protocol.

\subsection{Participants}
Eight secondary computing teachers took part in the PD workshop. Seven taught
at the high-school level, and one taught grades 6-8. Their recent
courses included introductory and advanced Computer Science, Python,
AP Computer Science Principles, Game Design, and drone technology. All teachers reported
holding a teaching license or another
state-recognized education credential in fields related to computing,
technology, business, or digital media. Table
\ref{tab:teachers} summarizes teachers' teaching experience and courses taught.

Teachers brought experience with a range of computing curricula
and platforms. All eight teachers reported prior use of at least one computing
curriculum or platform. Code.org was the most commonly reported resource
($n=8$), followed by CodeHS ($n=5$) and Scratch ($n=5$). Other resources
included Beauty and Joy of Computing, MIT App Inventor, CodeCombat, and Coding in Minecraft. Alongside these resources, all teachers reported using AI in some capacity during lesson preparation, with reported uses spanning initial ideation, content generation, and revision of existing materials.

\begin{table*}[t]
\caption{Participant's teaching experience}
\label{tab:teachers}
\small
\begin{tabularx}{\textwidth}{@{}l l X l@{}}
\toprule
\textbf{ID} & \textbf{Grades} & \textbf{Courses taught}
& \textbf{Teaching experience (years)} \\
\midrule

T01 & 9--12
& Computer Science Discoveries
& 16 \\

T02 & 9--12
& Introduction to Computer Science; Python I; Python II
& 5 \\

T03 & 11--12
& Digital media, Visual design, CTE 
& 21 \\

T04 & 9--12
& AP Computer Science Applications, Game Design
& 7 \\

T05 & 9--12
& Computer Science I, Fundamentals of Drone Technology
& 11 \\

T06 & 9--12
& Python I and Python II
& 21 \\

T07 & 9--12
& AP Computer Science I, 3D Modeling, Game Design
& 11 \\

T08 & 6--8
& Computer Science Discoveries, Coding in Minecraft
& 22 \\
\bottomrule
\end{tabularx}
\end{table*}

\subsection{Study 1: Semi-Structured Focus Groups}
\label{sec:focus-groups}

Study~1 examined teachers' existing lesson-authoring and adaptation
practices, their use and experience of generative AI, their
expectations for instructional control, and the criteria they use to
determine whether materials are ready for classroom use. We conducted
four semi-structured focus group interviews via Zoom, each included two
teachers and one moderator. Teachers were grouped by similarity in the
courses they taught. All the sessions lasted approximately 50-60 minutes and
were moderated by graduate researchers experienced in conducting
qualitative interviews and focus groups. 

Each session followed a common protocol organized into four modules
(Table~\ref{tab:fg-protocol}). Each module included a set of main
questions, optional follow-up probes, and a coverage checkpoint
specifying what the moderator needed to have elicited before moving to

\begin{table}[t]
\caption{Focus group protocol}
\label{tab:fg-protocol}
\small
\begin{tabular}{@{}p{0.42\columnwidth}p{0.52\columnwidth}@{}}
\toprule
\textbf{Modules} & \textbf{Topics and probes} \\
\midrule
\textbf{1. Lesson-Planning \& Authoring Workflow} & Starting points, resources used, stages, and revision or iteration points. \\
\textbf{2. Adaptation Needs \& Breakdowns} & Limitations of existing materials, adaptations needs and constraints \\
\textbf{3. Current AI Use, Limitations, \& Desired Control} & Experience with AI, its limitations, desired AI support and controls. \\
\textbf{4. Classroom Readiness \& Quality Review} & Criteria for determining classroom readiness, teacher judgments. \\
\bottomrule
\end{tabular}
\end{table}

\subsubsection{Focus Group Data Analysis}

We analyzed the transcripts from all four focus groups (eight
teachers) using a hybrid deductive-inductive thematic
analysis~\cite{fereday_demonstrating_2006}. The protocol's four
modules provided the initial deductive code families, while
inductive coding allowed us to identify additional distinctions
within teachers' accounts. Two researchers experienced in thematic
analysis first reviewed all transcripts independently, developing
a running list of subcodes as they encountered relevant ideas and
experiences in the data. They then met regularly to compare and
refine these subcodes, revise their organization within the code
families, and finalize the codebook. Both researchers subsequently
applied the codebook independently to the full corpus and resolved
coding disagreements through discussion.

Through this process, the initial organization was refined into
five final code families: (1) teachers' authoring starting points,
(2) adaptation for classroom fit, (3) selective AI delegation,
(4) AI breakdowns, and (5) classroom readiness. Once coding was
complete, we organized the coded data by teacher and code to
examine patterns, contrasts, and relationships across participants.
This comparison informed the findings reported in
Section~\ref{sec:focus-group-findings}.

\subsection{Study 2: Think-Aloud Tool Interaction}
\label{sec:tool-study}

Study~1 asked teachers to describe their existing lesson-preparation
practices, their use of AI within those practices, and the control
they exercised or sought over AI. These self-reported accounts
primarily concerned general-purpose conversational AI tools.
To complement these self-reported accounts, Study~2 used LessonBuilder as a research probe to examine how teachers specified instructional intent, directed AI actions, and exercised control while completing authoring tasks. LessonBuilder's explicit stages and inspectable
intermediate artifacts provided opportunities to observe these
decisions as teachers worked through the authoring process.

The study included two activities corresponding to LessonBuilder's
two authoring pathways
(Section~\ref{sec:authoring-workflows}):
(1) generating a new lesson on conditionals and
(2) importing and adapting existing materials for a lesson on loops.
To provide a common task context for examining teachers' decisions,
the topic for each activity was predetermined and held constant
across participants. At the beginning of each session, a researcher
provided a 10-minute demonstration of both workflows and invited
teachers to ask questions about the tool. Each of the eight teachers
then completed both activities individually in a moderated
think-aloud session. All sessions were recorded and transcribed.
The specific tasks included in each activity are described below.

\subsubsection{Activity 1: Generating and Revising a New Lesson}

In this activity, teachers were asked to create a lesson on
conditionals using LessonBuilder's generation pathway. The task
included the following operations, with opportunities to revisit
earlier stages and refine the plan and materials:

\begin{enumerate}
\item \textbf{Specify the lesson request:}
Teachers provided a prompt for a lesson on conditionals,
specifying instructional requirements such as grade level and
instructional style to guide the initial lesson plan.

\item \textbf{Inspect and revise the lesson plan:}
Teachers reviewed the proposed topics, timing, and activity types,
along with questions posed by the Planner to clarify or refine
the lesson requirements. They could revise the plan iteratively
before proceeding to material generation.

\item \textbf{Generate and inspect lesson materials:}
Teachers requested material generation from the revised lesson
plan and inspected the resulting learning content. They could
return to earlier stages for further revisions when the materials
did not meet their instructional requirements.

\item \textbf{Review AI quality feedback:}
Teachers requested an AI Quality Review to identify potential
issues in the generated materials relative to their instructional
requirements, providing additional review guidelines when needed.
Teachers considered the identified issues and suggested improvements
and revised the materials if they deemed changes necessary.

\item \textbf{Publish the materials:}
After inspecting the materials and making any further revisions,
teachers published the lesson in the application.

\end{enumerate}

\subsubsection{Activity 2: Importing and Adapting Existing Materials}

In this activity, teachers were asked to adapt a lesson on loops
sourced from a public forum for high-school teachers. Before the
study, the lesson was reviewed for suitability by a senior
high-school computing teacher with more than 10 years of experience
teaching computing courses. This teacher was not part of the study
cohort. The task included the following operations:

\begin{enumerate}
\item \textbf{Upload the lesson document:}
Teachers uploaded the provided document to obtain an import
preview containing the system's proposed topics and items.

\item \textbf{Inspect and edit the import preview:}
Teachers reviewed and edited the proposed topics and items before
confirming their addition to the class.

\item \textbf{Adapt the imported materials with AI assistance:}
Following publication, teachers used the Teaching Assistant to
revise the imported lesson materials.
\end{enumerate}

During both think-aloud activities, moderators did not suggest
what teachers should add, remove, or revise, and intervened only
to clarify confusion about the interface. Within the assigned
tasks, teachers retained responsibility for instructional decisions,
including which adaptations to make. Because both activities
required teachers to perform particular operations, completion
of an operation was not treated as evidence of an authoring
preference or control requirement. Instead, the analysis focused
on teachers' stated intentions and rationales, their choices
about what to change or retain, the revisions they made, their
evaluations of generated or imported materials, and their
explanations of where the system supported or constrained
their intended work.


Following the activities, teachers participated in a structured
post-activity reflection covering three areas: how generating a
new lesson and adapting an uploaded lesson fit into their usual
preparation workflows; how the system supported or constrained
their agency and control when adapting materials after publication
across both activities; and which missing control, capability,
or quality check they considered most important for classroom
readiness. Study~2 thus collected think-aloud transcripts, screen
recordings, system interaction logs, lesson artifacts authored
by teachers, and post-activity reflection discussions.


\subsubsection{Think-Aloud and Post-Study Reflection Analysis}

We analyzed the think-aloud transcripts and post-study reflection
discussions using inductive coding~\cite{clarke_teaching_2013}.
Two researchers independently reviewed all transcripts and recorded
emerging codes and subcodes. For the think-aloud data, screen
recordings were consulted when teachers' verbal comments did not
clearly indicate what they were viewing or doing. The researchers
met regularly to compare and refine their codes and organize
patterns concerning teachers' intentions, actions, evaluations,
and system behavior into four categories:
\emph{supported instructional intent}, when the system supported
teachers' intended work; \emph{additional authoring requirements},
when the data revealed needs not adequately supported by the
system; \emph{constrained instructional intent}, when system
behavior constrained, obscured, or altered teachers' intentions;
and \emph{anticipated use in existing practices}, when teachers
situated the system within their usual preparation practices.

Using the resulting codebook, both researchers independently
coded the full Study~2 qualitative corpus and resolved coding
differences through discussion. The same categories were applied
to the post-study reflections because these discussions also
addressed instructional intent, control, adaptation, and classroom
use. The think-aloud data captured teachers' actions and comments while
using LessonBuilder, while the post-study reflections captured
their assessments of the experience, anticipated uses in their
usual practice, and requests for additional support. We report the resulting findings in
Section~\ref{sec:study2-results}.


%% file: 05_results_S1.tex
\section{Results}
\label{sec:results1}

\subsection{Study 1: Reported Authoring and Adaptation Practices}
\label{sec:focus-group-findings}

To address RQ1, we examined how secondary computing teachers
created and adapted lesson materials for their classroom needs,
including the roles of AI and teacher judgment in this work.
Our analysis of the focus-group transcripts identified five
recurrent findings concerning how teachers prepared, adapted,
and evaluated lesson materials.
Table~\ref{tab:fg-findings} summarizes participant coverage for
each finding. Each count represents the number of teachers who
provided at least one substantive statement or example supporting
that finding in the coded transcripts, including the evidence not listed in this section. These counts describe coverage
within our sample; they do not represent the number of mentions
or the prevalence of these practices among teachers more generally.

\begin{table}[t]
\caption{Focus-group findings and participant coverage. Counts represent
teachers who provided substantive evidence for each finding in the
coded transcripts, including evidence not quoted in the text.}
\label{tab:fg-findings}
\small
\renewcommand{\arraystretch}{1.25}
\setlength{\tabcolsep}{4pt}

\begin{tabular}{
@{}
p{0.66\columnwidth}
p{0.28\columnwidth}
@{}
}
\toprule
\textbf{Finding} &
\textbf{Participant coverage} \\
\midrule

Authoring Starting Points
& 8/8 (T01--T08) \\

Adapting for Classroom Fit
& 8/8 (T01--T08) \\

Selective AI Delegation
& 8/8 (T01--T08) \\

AI Output Breakdowns and Repair
& 4/8 (T02, T03, T06, T07) \\

Judging Classroom Readiness
& 7/8 (T01, T03--T08) \\

\bottomrule
\end{tabular}
\end{table}

\subsubsection{Authoring Starting Points}
\label{sec:res1_1}
All eight teachers described beginning lesson preparation with
existing sources rather than creating a lesson entirely anew. These
sources including standards, objectives, curriculum
guides, course-specific resources, and materials they had
previously created or used.

Six teachers (T01, T02, T03, T05, T06, and T07) explicitly described
consulting standards, objectives, or curriculum guides when identifying
lesson goals and content. T05 described beginning with the intended
outcome and working backward: ``I start with the goal, and then I look to see what I would have to do to meet the goal.'' T03 similarly explained that the first thing I do is look at the standard objectives,
and then, the guide as well.'' T01 also considered the learning outcomes and
vocabulary students should know along with standards and objectives, while T02 examined district-provided
lessons in relation to the applicable standard and topic.

Course-specific resources also provided starting points.
T08 followed Code.org's Computer Science Discoveries curriculum
``almost step-by-step'' for a middle-school course, while T04
used its Computer Science Principles materials. T02 and T06
relied on CodeHS, and T05 described a statewide Moodle repository
where CTE teachers exchanged materials. When these resources
did not meet their needs, teachers sought additional support.
T02 consulted a district professional learning community,
instructional videos, and technical websites. T06 contacted
another district teacher after struggling to find materials
for teaching flowcharts in a Python course.

Teachers also revisited their own materials. T03 reused and
improved earlier materials, while T07 revised lessons from
previous years based on what had and had not worked during
instruction. 

\subsubsection{Adapting for Classroom Fit}

Established curricular requirements and resources did not ensure
that materials fit every classroom. Teachers adapted them to
address coverage, pacing, prior knowledge, reading level, and
class composition. As T04 observed, ``the curriculum doesn't
change that much, but the students do.'' T08 similarly described
reshaping materials to fit both the teacher and the students.

Teachers supplemented and updated materials to address gaps
in existing curricula and assignments.
T05 described a one-semester computer science curriculum spread
across Computer Science~1 and Computer Science~2 over an academic
year: ``The material is still good, it's just not enough.''
T05 and T01 supplemented their curricula with examples and
practice materials. T02 updated programming assignments to keep
them current and to make it more suitable for student portfolios.
T06 sought new activities when students did not relate to
existing materials and varied activities within 90-minute
classes to sustain engagement.

Teachers also differentiated materials according to learners'
needs. T06 considered students' age, community context, and
reading level, noting that ``a lot of kids are coming in maybe
sixth-, seventh-grade reading level.'' T03 created beginner,
standard, and advanced worksheets for the same project by
varying the language of instructions and the number of steps.
T07 likewise simplified directions for individual beginners.

Adaptation also involved changing how content was presented
and how students participated. To address students' divided
attention, T04 shortened explanations and incorporated physical
activities, discussion prompts, and peer work. For example,
T04 expanded a small Code.org exercise into a physical
demonstration of functions and parameters to make a difficult
concept easier to teach. T02 similarly favored shorter or
visual presentations over lengthy text: ``if we can chunk
things or present things graphically, it helps.''

These changes continued during and after classroom use.
T08 revised wording, timing, and activity placement across
three classes taught on the same day, explaining that
``by the third class \ldots{} I've adapted it and have
changed it.'' T07 revisited lessons based on what had and
had not worked, while T02 explained:
``I might have to come up with a worksheet between first
and third period.'' The extent of adaptation nevertheless
varied with course and institutional requirements. T05
reported relatively few student-specific changes beyond
accommodations, citing a flipped-classroom approach and
district-wide guidelines, while still adding supplementary
materials for coverage and differentiation.

\subsubsection{Selective AI Delegation}

Teachers used AI for specific lesson-preparation tasks,
including brainstorming, creating additional materials, and
adapting existing content. Being able to complete these tasks
quickly was a benefit explicitly described by four teachers
(T02, T03, T04, and T06). T02 used AI to create games,
worksheets, and small-group activities, explaining that
finding such materials between classes had previously taken
considerable time: ``with AI, it makes our life a lot more
doable.'' T03 described being able to generate slides from
an existing project within minutes, while T06 obtained
activities for lessons to be taught within the next day
or two. Other tasks included supplementing an incomplete
curriculum (T01) and converting lecture notes into charts
and other visual representations (T06).

To guide this work, teachers specified what they wanted
AI to do and the instructional requirements it should
consider. All eight described including the intended
change, instructional context, or both in their requests.
For example, T08 asked ChatGPT to reorganize a worksheet
from the Code.org curriculum into ``four workable stations.'' and adjusted
the proposed stations to fit the teacher and students
T04 requested a ``quick, short example'' to connect Code.org
and College Board materials, asking AI to incorporate
terminology such as \emph{procedure} for \emph{function}
and \emph{list} for \emph{array} into existing problems.
Teachers also described the learners for whom materials
were intended. T05 specified the topic, school level,
and students' existing knowledge, while T06 included
students' age, grade, community context, and reading level.

These requests also reflected limits on the work teachers
wanted AI to undertake. T05 generally requested supplementary
materials for differentiation rather than changes to existing
content. T03 provided standards, an outline, and previous
materials to guide the response: ``I don't depend on AI
fully. I would usually feed it the outline of the things
that I want. I just want it to organize it.'' Teachers
continued to evaluate and revise the results. T07 regarded an AI-generated lesson as ``a really
good idea'' that could be adapted for a particular class,
even when substantial revision was needed. 

\subsubsection{AI Output Breakdowns and Repair}

Teachers described difficulties getting AI to follow their
instructions and produce materials suitable for their classes.
Their accounts described three kinds of breakdown: materials
that did not fit learners' needs, changes to content teachers
intended to preserve, and unreliable information. Addressing
these problems required teachers to inspect the output,
clarify their requests, or revise the materials.

Mismatch with learners' needs was a recurring concern. T03
described evaluating AI-generated differentiated materials as
the most effortful part because ``even if you say, make it
simple, sometimes the product that AI gives isn't really what
we want.'' T02 encountered code beyond students' current
knowledge, ``including functions even though students didn't
even learn functions yet''. T02 also described making revision
requests more specific by asking for reading notes ``for very
beginner high school students in Python'' and specifying that
the notes should use only programming concepts students had
already learned. T07 similarly found that generated lessons
could be too difficult or too easy, requiring simplification,
extension, or substantial revision.

AI also changed content teachers wanted to preserve. T03
provided topics and activities planned for each day of the
week so that AI would organize the content, yet the system
would ``change it to something else.'' Clarifying a request
did not always resolve the mismatch: T02 described receiving
``the wrong thing over and over again'' and repeatedly
restating the request.

Other breakdowns concerned the reliability of generated
information. T02 received links to supporting videos or
webpages that did not exist. T06 recalled using an
AI-supported activity while rushed and discovering an
error during class. Warning that generated material might
``look smart'' while still being incorrect, T06 concluded
that ``we have to check it before we push it out.'' Checking and repairing these outputs could offset the
anticipated time savings. T07 reported having to ``spend
more time editing the lesson than it would to just have
created it in the first place.'' 

\subsubsection{Judging Classroom Readiness}

Correcting errors in AI-generated materials was only one part
of deciding whether a lesson was ready to teach. Teachers
remained responsible for that judgment, as T01 explained:
``I still have to physically teach the class. AI doesn't
teach the class.'' They considered whether the lesson
matched their instructional goals, provided sufficient
content, and was appropriate for their students. T05
summarized these considerations through three questions:
``Does it cover the objective? Is it rich in material?
And will my students be able to understand it or learn
it as it's written?'' T03 and T07 also considered
alignment with standards, language and vocabulary, and
students' existing knowledge. Cognitive demand mattered
as well: T06 explained, ``We don't want the kids just
to be recalling all the time,'' and described balancing
recall with higher-level thinking.

Teachers also considered whether activities would work
under actual classroom conditions. T04 described using
Duck Duck Goose to represent conditionals and emphasized
that an activity's feasibility became apparent through
classroom use. In a mixed-level middle-school class,
T08 assessed whether students understood enough to
participate before starting an activity and kept a
paper handout as an alternative if they struggled.

Timing was another aspect of classroom feasibility.
T07 recounted that a nominally 90-minute plan could
progress much more slowly during instruction, explaining
that ``AI can't tell me what my class is going to
accomplish in one period.'' T03 similarly noted that
AI could produce a ``beautifully made lesson plan with
the perfect timing,'' although an activity allocated
five minutes might take ten or fifteen depending on
the students that day. Across these accounts, classroom
readiness depended on whether the complete lesson could
meet its instructional goals with the students and
time available.


%% file: 05_results_S2.tex
\subsection{Study 2: Negotiating Instructional Intent and Control Through Staged Authoring}
\label{sec:study2-results}

\begin{table}[t]
\caption{Think-aloud tool interaction findings and participant coverage. Counts represent
teachers who provided substantive evidence for each finding in the
coded transcripts, including evidence not quoted in the text.}
\label{tab:fg-finding-s2}
\small
\renewcommand{\arraystretch}{1.25}
\setlength{\tabcolsep}{4pt}

\begin{tabular}{
@{}
p{0.66\columnwidth}
p{0.28\columnwidth}
@{}
}
\toprule
\textbf{Finding} &
\textbf{Participant coverage} \\
\midrule

Teachers Assessed Classroom Fit Across Planning and Review Phases
& 5/8 (T01, T02, T03, T05, T07) \\

Plan Review Surfaced Additional Instructional Decisions
& 3/8 (T03, T04, T08) \\

Quality Review Informed Rather Than Determined Teacher Judgment
& 3/8 (T03, T04, T07) \\

Post-Study Reflections
& 6/8 (T02, T03, T05, T06, T07, T08) \\

\\
\bottomrule
\end{tabular}
\end{table}

Addressing RQ2, Study~2 examined how secondary computing teachers negotiated instructional intent and control while using LessonBuilder's structured authoring process. Teachers often clarified or revised their intentions as they inspected successive outlines, lesson materials, revisions, and AI feedback. They also checked whether earlier decisions remained reflected and whether the system's outputs suited their students and curricula. Teacher control therefore continued throughout the authoring process rather than occurring only when teachers submitted the initial request or approved the final lesson.

The four categories described in the Methods guided how we organized the findings below. Evidence coded as \emph{supported instructional intent} and \emph{additional authoring requirements} informed the finding that teachers expressed lesson requirements in combinations. Evidence coded as \emph{constrained instructional intent} showed how those requirements were refined, preserved, or lost during revision and how teachers responded to AI quality feedback. Evidence coded as \emph{anticipated use in existing practices} informed the final finding about how LessonBuilder fit teachers' existing materials and preparation practices. These findings describe recurring patterns rather than fixed authoring steps.

\subsubsection{Teachers Assessed Classroom Fit Across Planning and Review Phases}
During lesson creation, teachers expressed instructional intent through both structured selections and natural-language requests. These requirements extended beyond topic and grade level to include lesson duration, students' prior knowledge, instructional sequencing, accessibility and visual supports, teacher-facing materials, standards, and learning objectives. However, the structured options did not always accommodate combinations of requirements. The instructional-style field permitted only one selection: T02 wanted to select ``several of these criteria,'' T03 wanted the lesson to be both beginner-friendly and project-based, and T05 wanted it to be both beginner-friendly and creative. Teachers therefore had to use the prompt to communicate instructional combinations that the structured fields could not fully represent.

The prompt allowed teachers to express pedagogical relationships beyond the structured options. T03 asked the AI planner for a guided, step-by-step exercise followed by an unguided exercise so that students could first develop understanding and then work independently. When both appeared in the proposed plan, T03 responded, ``It has the hands-on! Nice! Guided and independent!''

Reviewing the plan further helped teachers clarify how their requirements should be operationalized. T05 asked the AI planner to include both Java and Python because students began with Java but could later choose between the two languages. When the plan framed this requirement as a choice for students, T05 clarified, ``I wanted it to give me one of each.'' The visible plan therefore helped distinguish between supporting students' language choices and generating parallel versions of the materials.

The generated materials provided another point at which teachers assessed how a lesson would fit their instruction. T02 identified good vocabulary, a useful checkout-system scenario, and starter code, but situated the activity later in the course because it assumed that students already understood variables. T02 explained that beginning students would first work with strings, followed by integers and comparisons, and would need additional visual and preparatory materials. T02 consequently described the generated lesson ``as a base for me to start from.'' T07 similarly considered a generated lesson ``pretty close'' to the intended goals while identifying refinements to the introductory presentation, examples, practice activities, and review quiz.

Teachers also evaluated classroom fit against curricular knowledge beyond what was immediately visible in the generated artifact. T01 reported that changes to the outline appeared in the generated lesson and that the result remained broadly on task. Before approving it for classroom use, however, T01 would compare the lesson with the current objectives and with what students should know by the end of the unit. Together, these accounts show that teachers assessed and refined instructional intent across the prompt, proposed plan, and generated materials. These checkpoints supported teachers in making the system's interpretation visible while retaining responsibility for determining how the resulting materials fit their students, instruction, and curricula.

\subsubsection{Plan Review Surfaced Additional Instructional Decisions}

Teachers did not articulate every instructional decision in their initial requests. As they responded to the AI planner's questions and reviewed the evolving outline, they specified requirements that became salient during planning.

For T04, the planner's questions prompted consideration of whether a playground activity should use a fixed number of repetitions or randomness. Rather than choosing one, T04 developed a comparative activity in which different groups would use the two approaches to contrast for-loops with while-loops. T04 described these as ``great questions to make me think about it,'' adding, ``I've never thought about this before.'' The questions therefore prompted a new comparative structure for the activity rather than merely collecting missing information.

The planner's questions also elicited additional decisions about lesson content and implementation. When asked whether the lesson should introduce input or remain focused on variables, T03 chose to add input and then inspected how the updated outline represented that decision. T08 similarly revised the plan to use Scratch, account for students' prior experience with events and loops, situate variables in a collecting game, and organize the activity as pair work. Across these cases, plan review did more than confirm teachers' initial requests: it surfaced additional decisions about conceptual comparisons, prerequisites, activity formats, and learning environments before material generation.

\subsubsection{Quality Review Informed Rather Than Determined Teacher Judgment}

Teachers evaluated not only the generated lessons but also the AI reviewer's judgments. Rather than treating the Quality Review as definitive, they selectively accepted, revised, or contested its feedback using their classroom and curricular knowledge.

T03 valued the review for identifying concerns rather than simply affirming the requested lesson. It drew attention to duration, vocabulary, language, and grade level, including whether the planned activities could fit within the requested 60 minutes. While inspecting this feedback, T03 described ``the review as a guide for the teachers'' and explained that, ``unlike ChatGPT, it could indicate when a requested 60-minute plan doesn't look realistic.''

Other teachers qualified or contested particular review judgments. T07 disagreed with the estimated duration of approximately 40 minutes, judging the lesson closer to 45 minutes based on classroom experience, while concluding that ``everything else looks pretty good.'' T04 similarly questioned the review's assertion that program
termination was covered in AP Computer Science Principles, responding,
``I don't know about that termination thing'' and ``I haven't seen it.'' However, T04 accepted the observation that the generated content used Python rather than the JavaScript taught in the course and attributed this mismatch to not having specified JavaScript in the prompt.

Across these responses, the Quality Review surfaced issues for consideration without displacing teachers' judgment. Teachers determined which feedback was credible and relevant, leaving
the teacher, rather than the AI reviewer, as the final evaluator of
classroom readiness.

\subsubsection{Post-Study Reflections}

After completing both activities, teachers responded to three structured reflection questions concerning how generation and adaptation would fit their usual preparation, where the system supported or constrained the adaptations they typically made, and which missing control, capability, or quality check they considered most important for classroom use.

Teachers situated generation and adaptation according to the materials already available to them. T02 distinguished the generated lesson as the ``bones'' of an artifact from the more structured imported material, which appeared as ``polished content ready to deliver.'' T03 and T08 anticipated using adaptation more often because they generally began with available or provided materials. Generation nevertheless remained relevant when those materials did not address a particular need: T03 anticipated using it for new or differentiated content, while T05 described using AI-generated materials to supplement an existing scripted curriculum for students progressing at different rates. T07 argued for using the two pathways, explaining, ``I would, but then I'd want to be able to adapt it.'' These responses positioned generation and adaptation as complementary starting points rather than mutually exclusive workflows.

Choosing a starting point did not capture the full extent of the control teachers wanted. Their reflections emphasized the ability to direct and inspect particular changes after materials entered either workflow. T05 prioritized having AI revisions follow the teacher's request precisely, explaining, ``If I say this, this is what I want you to give me back, not you give me back what you want me to have.'' T07 wanted revision controls embedded within individual lesson sections so that a targeted component could be changed without leaving its editing context. In an additional follow-up, T02 and T07 supported extending the Quality Review to adaptation. T02 associated this capability with greater control over modifying an imported lesson, while T07 valued being able to inspect the adapted material quickly.

Teachers also identified controls that would extend the content and contexts the system could support. T02 prioritized specifying standards and learning objectives, reinforcing requirements raised during lesson creation by identifying them as the most important missing control for classroom use. T08 wanted to add assessment questions while preserving suitable existing items and proposed uploading standards so that they could ``play a part in the lesson.'' T08 also explained that the Python-oriented materials encountered during the study did not match the middle-school course taught, indicating that programming-language and course-level coverage would shape where the system could be used. T06 pointed to teacher-facing guidance and answer keys available in existing coding platforms as particularly helpful when teachers themselves needed support understanding or explaining programming material.

Across the activity episodes and post-study reflections, teachers
continued to exercise control as materials moved through prompts,
plans, generated or adapted artifacts, and AI feedback. They clarified
requirements, inspected how those requirements were represented, and
used their pedagogical and curricular knowledge to determine which
outputs and recommendations were appropriate. Generation and
adaptation provided different starting points, but both required
opportunities for targeted revision and teacher-led evaluation before
materials could be incorporated into classroom preparation.

%% file: 06_framework.tex
\section{Cross-Study Synthesis: Conceptual Model and Design Recommendations}

\label{sec:cm}
\begin{figure*}[t]
\centering
\includegraphics[width=\textwidth]
{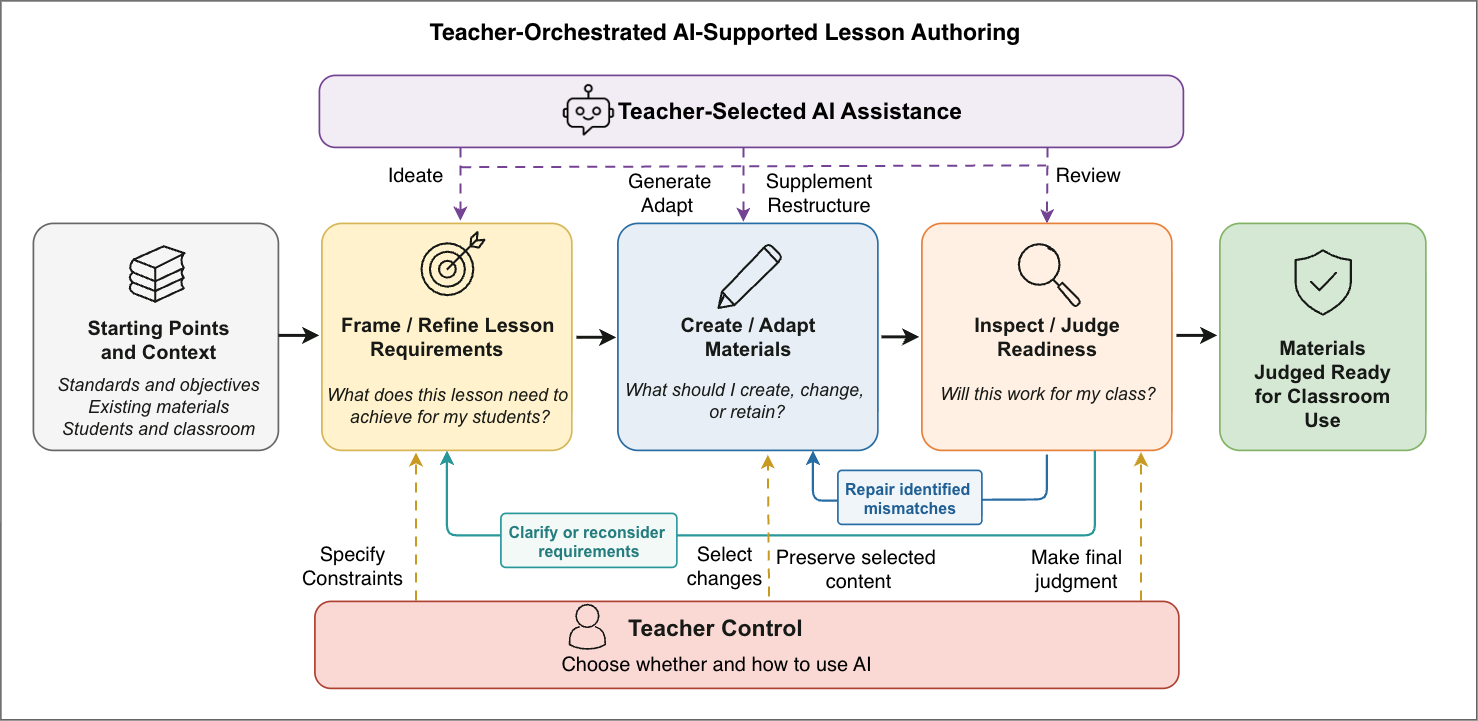}
\caption{Conceptual model of teacher-orchestrated AI-supported
lesson authoring: teachers begin from \colorbox{modelGray!25}{Starting points and context}
\colorbox{modelGold!15}{frame requirements},
\colorbox{modelBlue!20}{create or adapt materials}, and
\colorbox{modelOrange!10}{judge readiness}, with
\colorbox{modelPurple!25}{teacher-selected AI assistance} and
\colorbox{modelRed!25}{teacher control} throughout.
Solid arrows show possible transitions and revision loops;
dashed arrows connect AI assistance and teacher control
to the activities.}
\label{fig:conceptual-model}
\end{figure*}
The focus groups and tool study together suggest that computing
lesson authoring involves revisiting instructional requirements
as materials take shape. Teachers refine what a lesson should
accomplish while creating, adapting, and reviewing its content.
This relationship between requirements and developing materials
is consistent with Cui et al.'s conceptual model of assessment
authoring
\cite{cui_codesigning_2026}.

Our cross-study synthesis examines this relationship in computing
lesson authoring, focusing on how teachers decide which tasks to
delegate to AI, which instructional choices to preserve, and how
to respond when outputs do not meet their requirements. It also
considers how teachers judge whether the resulting materials are
ready for a particular classroom. We organize these relationships
in a conceptual model of teacher-orchestrated AI-supported lesson
authoring (Figure~\ref{fig:conceptual-model}), where teachers
collaborate with AI while retaining responsibility for their
instructional decisions. The model represents recurring activities
and decisions rather than a fixed sequence, and informs the design
recommendations for future AI-orchestrated systems for teacher lesson authoring.

\subsection{Teacher-Orchestrated AI-Supported Lesson Authoring}

Teachers engage in an \emph{iterative authoring process} that begins from an initial starting point, moves through framing or refining instructional needs, creating or adapting materials, and inspecting classroom readiness, it ends when materials are judged ready for use. Second, throughout this authoring process teachers may \colorbox{modelPurple!25}{\emph{delegate selected work to AI}} rather than transferring the authoring process itself to AI. Third, teachers \colorbox{modelRed!25}{\emph{retain instructional control}} by making decisions, revising developing materials, and ultimately judging the lesson's classroom readiness. Together, these components characterize AI-supported lesson authoring as a form of human–AI collaboration in which teachers coordinate AI contributions within an iterative authoring process while retaining control over instructional intent, revision, and classroom-readiness decisions.

\paragraph{\textbf{Iterative authoring process}} Teachers prepare their lesson from a \colorbox{modelGray!25}{starting point}, which includes existing or external materials, such as curricula, lessons, examples, assignments, and instructional platforms. From this starting point, teachers first \colorbox{modelGold!15}{\emph{frame their instructional needs}}, specifying what students should learn, practice, and demonstrate. Such needs draw on curricular goals, students' current understanding, instructional time and context. Teachers make
these requirements explicit through details such as the
computing topic, grade level, students' prior knowledge,
lesson duration, instructional style, and desired activities.

Although clearly formulated needs give teachers a basis for creating or adapting materials, intent can remain incomplete, and teachers may return to this phase after inspection when a generated or an existing lesson reveals a missing objective, an incorrect learner level, insufficient support, or an unrealistic pacing assumption.

In the next step, teachers \colorbox{modelBlue!25}{\emph{create or materials}} by combining resources from curricula, colleagues, online sources, and their own previous lessons. Teachers
combine these resources by preserving useful content
while changing what does not fit the current class. In
computing lessons, this work includes developing or revising
explanations, programming examples, practice activities,
assignments, and assessment questions.

Once a draft or revision is available, teachers \colorbox{modelOrange!10}{\emph{inspect the materials
and judge their classroom readiness}}. Teachers check whether materials aligned with their objectives, matched students' knowledge and needs, fit the available time, and could work under classroom conditions. For computing content, this
also involves examining whether examples are clear,
programming activities are appropriately challenging,
and technical information is accurate and suitable for
the learners.

Inspection can lead teachers back to different parts of
the authoring process. A draft may reveal that a requirement
needs clarification or reconsideration, such as an overlooked
learning objective, a need for additional support, or an
unrealistic pacing expectation. Alternatively, the requirement
may remain appropriate while the material fails to meet it;
in that case, teachers revise the material while preserving
the original requirement. Teachers may move between these
activities until they judge the lesson ready for classroom
use. This judgment reflects the intended class and context,
does not guarantee pedagogical effectiveness or technical
accuracy, and may be reconsidered during teaching.

\paragraph{\colorbox{modelPurple!25}{\textbf{Delegate bounded work to AI}}}
Teachers delegate their work to AI in multiple stages of this process. When specifying and refining needs, teachers adopt AI for ideation and creation of lesson plans. When creating or adapting materials, teachers use AI to generate, supplement, adapt, or restructure materials, producing learning activities, examples and questions, revising generated materials, and restructuring content. When judging the lesson's readiness, teachers refer to AI review to as a source of information for decision making. 

Here, delegation is bounded because teachers invoke AI for particular authoring tasks with clear boundaries rather than handing over the lesson-authoring process as a whole. AI contributes specific materials, transformations, or suggestions at different points in the process, with its role determined by the work teachers choose to delegate. Its outputs then return to the broader authoring process for further inspection and possible refinement rather than directly determining that the lesson is complete or classroom ready.

\paragraph{\colorbox{modelRed!25}{\textbf{Retaining teacher control}}} Despite the involvement of AI, teachers retain critical instructional control throughout the process. This control defines the boundaries around AI involvement: teachers specify constraints that AI should respect, determine which parts of a lesson may change and which should remain unchanged, retain the ability to make changes manually, and decide whether AI-generated materials or recommendations should be incorporated.

These forms of control appear across each stage of the authoring process. When framing or refining needs, teachers establish the instructional constraints that subsequent work should preserve, including learning goals, learner needs, pacing, and instructional approaches. When creating or adapting materials, teachers exercise selective control over the scope of change by identifying what should be modified while preserving content they already consider appropriate. When inspecting and judging readiness, teachers retain final authority to accept, revise, or reject AI-produced materials and recommendations and to determine whether the resulting lesson is ready for classroom use. However, retaining this authority does not mean that AI
consistently follows teachers' instructions. Outputs may
change content teachers intended to preserve or fail to
meet a stated requirement. Exercising control therefore
also includes the work needed to address these mismatches:
restating a request, revising materials manually, or
returning to an earlier authoring activity. Our model
captures both the decisions teachers retain and the effort
that may be required to carry those decisions through
AI-supported authoring.

\subsection{Design Recommendations for AI-Supported Lesson Authoring}
\label{sec:design-recommendations}

Drawing on the conceptual model and findings from both studies,
we propose four design recommendations for AI-supported lesson
authoring. These recommendations address how systems can support
selective revision, evolving instructional requirements, and
teacher judgment, alongside the particular demands of computing
materials.

\paragraph{\textbf{DR1: Support selective revision}}
Our model indicates that teachers may delegate AI in bounded tasks, such as adding an example, while intending other instructional decisions to remain unchanged. Our findings further show that AI can violate these boundaries by altering content or assumptions that teachers intended to preserve, creating additional repair work. Systems should therefore represent not only the target and requested transformation, but also the constraints or lesson elements that should remain unchanged. When these constraints are violated, teachers should be able to identify the affected component, make local edits, or return to an earlier authoring state without regenerating unaffected materials. This extends prior notions of \emph{agency}: systems should not only allow teachers to initiate or approve AI actions, but also allow them to define the boundaries within which those actions may alter the lesson.

\paragraph{\textbf{DR2: Track evolving lesson requirements}}
Our model characterizes lesson intent as something teachers repeatedly refine rather than fully specify at the beginning. During staged authoring, reviewing intermediate plans can surface additional intents that were not completely expressed in the initial request. Systems should therefore maintain an explicit and revisable representation of instructional intent across the authoring process, including both the initial intents and intents established during earlier revisions. Intermediate artifacts can then help teachers compare materials against this evolving intent and revise when mismatches emerge. This extends prior accounts of \emph{transparency}: rather than making only AI outputs or intermediate states visible, system should also make the current representation of the teacher's instructional intent inspectable and revisable as that intent develops. This recommendation builds on
work connecting evolving requirements with reusable authoring
operations \cite{cui_codesigning_2026}.

\paragraph{\textbf{DR3: Support teacher judgment of AI feedback}}
Teachers evaluated materials using curricular and classroom
knowledge that was not fully represented in the system.
During the tool study, they accepted some AI review observations
while questioning or rejecting others. This suggests that,
when authoring systems provide AI-generated evaluations or
revision advice, teachers need support for determining whether
that feedback applies to their classroom. Systems should connect
each concern to the relevant material and explain the criterion
or assumption behind it. Teachers should be able to question
these assumptions, request reassessment, and act on individual
suggestions without accepting the entire review. Such support
would help teachers use AI feedback as one source of information
while retaining responsibility for judging classroom readiness.

\paragraph{\textbf{DR4: Provide computing-specific support for authoring}}
The model and our other findings show that teachers in computing education create, adapt, and inspect materials that are unique to the field of computing, such as programming examples, with specific requirements including difficulty of programming activities, technical accuracy and feasibility. Computing lesson-authoring systems should therefore treat programming examples, executable code, starter code, solutions, test cases, and technical environments as first-class authoring components. Systems could support operations such as code execution, test-case validation, programming-language or tool configuration, and inspection of technical dependencies or environment requirements. These capabilities would allow teachers to evaluate not only whether a lesson is pedagogically appropriate, but also whether its computing components can be executed, debugged, and used under the technical conditions of the classroom.

%% file: 07_discussion.tex
\section{Discussion}
\label{sec:discussion}

Across the two studies, teacher's adoption of AI was not limited to writing an initial prompt or approving a completed lesson. Teachers exercised control by establishing lesson requirements, deciding what material needed to change, selecting when to use AI, inspecting intermediate artifacts, and judging whether the result fit their classroom. We discuss these findings in relation to each research question.

\subsection{RQ1: Creating and Adapting Materials for Classroom Fit}

Our first research question examined how secondary computing teachers create and adapt lesson materials and what shapes this work. Teachers rarely described preparation as creating a complete lesson from an empty page. Instead, they began with standards, curriculum platforms, prior lessons, shared resources, or materials they had developed previously. They combined these sources and created additional content when available materials did not provide sufficient coverage. This supports prior accounts of teachers as designers who select, interpret, and adapt curriculum resources rather than simply implement them \cite{kaufman_how_2020,sherin_curriculum_2009,choppin_role_2022}.

A resource’s availability did not establish its suitability. Teachers adapted explanations, examples, assignments, pacing, activity formats, and levels of support to fit particular students and classroom conditions. Their readiness judgments connected curricular alignment with prior knowledge, reading level, engagement, available time, and classroom feasibility. Classroom readiness was therefore relational: the same material could be suitable for one class but require substantial change for another.

Computing added further dimensions to this judgment. Teachers considered programming language, conceptual prerequisites, the difficulty of programming activities, the accuracy of technical material, and whether students could use the required technologies. These concerns complement prior work showing that computing materials must account for technical correctness, varied programming experience, and the different errors and solution paths that programming activities can produce \cite{vinnervik_-depth_2023,friend_solve_2022,gale_hands-up_2026}. Supporting computing lesson preparation therefore requires more than producing pedagogically plausible text; systems must allow teachers to inspect and adapt the technical and instructional components of a lesson.

The extent of this work also depended on teachers' curricular responsibilities. Some teachers followed prescribed curricula closely and primarily supplemented or localized existing materials. Others regularly combined resources or developed new lessons. The amount of freedom to author materials should therefore be treated as a contextual condition. A system should support both extensive lesson construction and smaller adaptations without assuming that every teacher needs to generate a complete lesson.

\subsection{RQ2: Instructional Intent and Control with Structured AI Support}

Our second research question examined how teachers negotiated instructional intent and control while using LessonBuilder's structured workflow. The findings show that instructional intent was not always completely specified in the initial request. Teachers clarified what they wanted while responding to planning questions and inspecting proposed plans. Plan review surfaced decisions about prerequisites, programming languages, activity structure, collaboration, scaffolding, and conceptual comparisons before materials were generated.

The proposed plan therefore served as more than an intermediate system output. It provided a representation against which teachers could examine the system's interpretation and refine their own requirements. This finding supports staged authoring approaches that expose intermediate representations \cite{fan_lessonplanner_2024,varuvel_dennison_shiksha_2026,kaputa_simstep_2026}, while showing that their value is not limited to detecting generation errors. Intermediate artifacts can also help teachers articulate instructional decisions that become apparent only as a lesson develops.

Teachers continued this alignment after material generation. They compared explanations, programming activities, examples, quizzes, and time estimates with earlier decisions and with knowledge not fully represented in the system. Teacher control was therefore exercised through repeated comparisons between instructional intent and developing artifacts. This suggests that transparency is useful only when it is actionable. Showing an intermediate artifact provides limited control if teachers cannot identify what changed, revise the relevant component, or preserve material they have already accepted \cite{amershi_guidelines_2019,pareek_sensemaking_2026}.

The Quality Review created another inspection point, but it did not transfer evaluative authority to AI. Teachers considered its observations, accepted some, and contested others using their curricular and classroom knowledge. The AI review was itself an artifact requiring teacher evaluation rather than an independent determination of lesson quality. This distinction is important because an additional AI review does not verify factual correctness, pedagogical quality, or classroom readiness. It can direct attention to possible concerns, but the teacher determines whether those concerns are credible and relevant.

\subsection{RQ3: Teacher-Orchestrated AI-Supported Lesson Authoring}

Our third research question asked how the two studies inform a conceptual model of teacher control and corresponding design requirements. The resulting model characterizes lesson authoring as movement among three recurring forms of work: framing and refining intent, creating and adapting materials, and inspecting classroom readiness, with AI involvement throughout the process and teachers retaining control when collaborating with AI. It builds on prior work describing authoring as an iterative relationship between developing requirements and developing materials \cite{cui_codesigning_2026}. Our findings extend this account to computing lessons and make the placement of AI assistance explicit.

The model's central distinction is between deciding what transformation is needed and deciding who should perform it and how. Adding an example, supplementing practice, simplifying an explanation, or restructuring an imported lesson identifies the desired change. The teacher may make that change manually, delegate it to AI, or combine both approaches. AI delegation is therefore conditional rather than a required step in lesson authoring. This differs from models in which AI generation defines the workflow and teachers are primarily asked to refine or approve its output.

Bounded delegation also provides a more concrete account of teacher authority. Authority depends on teachers being able to define the intended change, select its target, choose whether to use AI, inspect the proposed change, preserve unaffected material, and decide whether the result should be incorporated. A single interface control cannot provide all of these capabilities. An accept-or-reject button may preserve final approval while offering little control over how an unsuitable output can be repaired.

Moreover our model should not be interpreted as a universal sequence followed by every teacher. Teachers may begin with an existing resource, a need for new material, or a request for AI assistance. They may also return to an earlier phase when inspection reveals a missing requirement or unsuitable component. Its contribution is therefore not a prescribed workflow but an account of the forms of work and transitions that AI-supported authoring systems should accommodate. The design recommendations in
Section~\ref{sec:design-recommendations} address selective
revision, evolving lesson requirements, teacher judgment
of AI feedback, and technical and instructional inspection
of computing materials.

\section{Limitations}
\label{sec:limitations}

We examined secondary computing teachers' lesson-preparation
practices and their use of LessonBuilder to understand how they
direct AI assistance during authoring. The resulting conceptual
model and design recommendations should be considered in light
of the participant context, study activities, and duration of
tool use.

The research involved eight secondary computing teachers from
one professional development cohort. Although participants
brought varied teaching experience and curricular responsibilities,
recruitment within this cohort limited the range of educational
contexts represented. Teachers working under different curricular
requirements, institutional conditions, or levels of familiarity
with AI may have additional needs or exercise control differently.
Further research across these settings could examine which
aspects of the model transfer and which require adaptation.

The study activities also shaped how teachers' authoring decisions
became visible. Focus groups captured teachers' accounts of their
usual preparation, while the tool study allowed us to examine
their choices and explanations as they worked with LessonBuilder.
Predetermined topics and assigned operations provided a common
context for examining these decisions, but the interface and
instructions influenced the available actions and their sequence.
The observed patterns therefore reflect both teachers' responses
to the authoring tasks and the opportunities provided by the
system. The conceptual model organizes recurring forms of
authoring work and teacher control; the sequences observed during
the sessions should not be interpreted as teachers' preferred
order of activities in everyday practice.

These conditions also limit what we can infer about how
AI-supported authoring develops over time. Each teacher used
LessonBuilder within a single session, providing insight into
initial interactions rather than sustained use. As teachers
become familiar with the system, they may change what they
delegate, how they specify requirements, and how they inspect
AI outputs. Moreover, classroom readiness was assessed during
authoring rather than through subsequent classroom use.
Following teachers and their materials over time would help
examine how these judgments hold up during instruction and
how classroom experience informs later revisions. Such work
would also allow the design recommendations to be evaluated
within ongoing lesson preparation.

%% file: 08_conclusion.tex
\section{Conclusion}

This work examined how secondary computing teachers prepare
and adapt lesson materials and exercise instructional control
when using AI support. Bringing together teachers' accounts
in focus groups and their interactions with LessonBuilder,
our findings highlight the iterative relationship among
refining instructional intent, developing materials, and
judging classroom readiness. Teachers selectively delegated
work to AI while expecting their instructional choices to
be preserved. Their responses to intermediate plans,
generated materials, and quality-review feedback illustrate
how teacher control involves inspecting and revising AI
contributions as well as deciding whether to accept them.

We synthesize these findings in a conceptual model of
teacher-orchestrated AI-supported lesson authoring. The
model distinguishes deciding what authoring work is needed
from deciding whether to delegate that work to AI, positioning
AI assistance as optional across recurring authoring activities.
The corresponding design recommendations address selective
revision, evolving lesson requirements, teacher judgment of
AI feedback, and technical and instructional inspection of
computing materials. Together, these contributions offer
a basis for designing systems that help teachers carry
their instructional decisions through AI-supported authoring
and retain control over what becomes part of a classroom lesson.

%% file: sigconf.bbl

\begin{thebibliography}{47}


\ifx \showCODEN    \undefined \def \showCODEN     #1{\unskip}     \fi
\ifx \showISBNx    \undefined \def \showISBNx     #1{\unskip}     \fi
\ifx \showISBNxiii \undefined \def \showISBNxiii  #1{\unskip}     \fi
\ifx \showISSN     \undefined \def \showISSN      #1{\unskip}     \fi
\ifx \showLCCN     \undefined \def \showLCCN      #1{\unskip}     \fi
\ifx \shownote     \undefined \def \shownote      #1{#1}          \fi
\ifx \showarticletitle \undefined \def \showarticletitle #1{#1}   \fi
\ifx \showURL      \undefined \def \showURL       {\relax}        \fi
\providecommand\bibfield[2]{#2}
\providecommand\bibinfo[2]{#2}
\providecommand\natexlab[1]{#1}
\providecommand\showeprint[2][]{arXiv:#2}

\bibitem[Albaloul et~al\mbox{.}(2026)]%
        {albaloul_systematic_2026}
\bibfield{author}{\bibinfo{person}{Omar Albaloul}, \bibinfo{person}{Abdulrahman~S. Alajmi}, \bibinfo{person}{Nicole Oster}, {and} \bibinfo{person}{Chad Killian}.} \bibinfo{year}{2026}\natexlab{}.
\newblock \showarticletitle{Systematic review of qualitative studies exploring {K}-12 teachers’ perceptions and experiences of using {ChatGPT}}.
\newblock \bibinfo{journal}{\emph{Discover Artificial Intelligence}} \bibinfo{volume}{6}, \bibinfo{number}{1} (\bibinfo{date}{June} \bibinfo{year}{2026}), \bibinfo{pages}{513}.
\newblock
\showISSN{2731-0809}
\href{https://doi.org/10.1007/s44163-026-01344-7}{doi:\nolinkurl{10.1007/s44163-026-01344-7}}


\bibitem[Amershi et~al\mbox{.}(2019)]%
        {amershi_guidelines_2019}
\bibfield{author}{\bibinfo{person}{Saleema Amershi}, \bibinfo{person}{Dan Weld}, \bibinfo{person}{Mihaela Vorvoreanu}, \bibinfo{person}{Adam Fourney}, \bibinfo{person}{Besmira Nushi}, \bibinfo{person}{Penny Collisson}, \bibinfo{person}{Jina Suh}, \bibinfo{person}{Shamsi Iqbal}, \bibinfo{person}{Paul~N. Bennett}, \bibinfo{person}{Kori Inkpen}, \bibinfo{person}{Jaime Teevan}, \bibinfo{person}{Ruth Kikin-Gil}, {and} \bibinfo{person}{Eric Horvitz}.} \bibinfo{year}{2019}\natexlab{}.
\newblock \showarticletitle{Guidelines for {Human}-{AI} {Interaction}}. In \bibinfo{booktitle}{\emph{Proceedings of the 2019 {CHI} {Conference} on {Human} {Factors} in {Computing} {Systems}}} \emph{(\bibinfo{series}{{CHI} '19})}. \bibinfo{publisher}{Association for Computing Machinery}, \bibinfo{address}{New York, NY, USA}, \bibinfo{pages}{1--13}.
\newblock
\showISBNx{978-1-4503-5970-2}
\href{https://doi.org/10.1145/3290605.3300233}{doi:\nolinkurl{10.1145/3290605.3300233}}


\bibitem[Calò et~al\mbox{.}(2026)]%
        {calo_cat-ai_2026}
\bibfield{author}{\bibinfo{person}{Tommaso Calò}, \bibinfo{person}{Lorenzo Cuccu}, {and} \bibinfo{person}{Luigi De~Russis}.} \bibinfo{year}{2026}\natexlab{}.
\newblock \showarticletitle{{CAT}-{AI}: {Supporting} {Teacher} {Workflows} with {AI}-{Assisted} {Exercise} {Creation}}. In \bibinfo{booktitle}{\emph{Proceedings of the {Extended} {Abstracts} of the 2026 {CHI} {Conference} on {Human} {Factors} in {Computing} {Systems}}} \emph{(\bibinfo{series}{{CHI} {EA} '26})}. \bibinfo{publisher}{Association for Computing Machinery}, \bibinfo{address}{New York, NY, USA}, \bibinfo{pages}{1--6}.
\newblock
\showISBNx{9798400722813}
\href{https://doi.org/10.1145/3772363.3799367}{doi:\nolinkurl{10.1145/3772363.3799367}}


\bibitem[Cheah et~al\mbox{.}(2025)]%
        {cheah_integrating_2025}
\bibfield{author}{\bibinfo{person}{Yin~Hong Cheah}, \bibinfo{person}{Jingru Lu}, {and} \bibinfo{person}{Juhee Kim}.} \bibinfo{year}{2025}\natexlab{}.
\newblock \showarticletitle{Integrating generative artificial intelligence in {K}-12 education: {Examining} teachers’ preparedness, practices, and barriers}.
\newblock \bibinfo{journal}{\emph{Computers and Education: Artificial Intelligence}}  \bibinfo{volume}{8} (\bibinfo{date}{June} \bibinfo{year}{2025}), \bibinfo{pages}{100363}.
\newblock
\showISSN{2666-920X}
\href{https://doi.org/10.1016/j.caeai.2025.100363}{doi:\nolinkurl{10.1016/j.caeai.2025.100363}}


\bibitem[Chiang et~al\mbox{.}(2023)]%
        {chiang_are_2023}
\bibfield{author}{\bibinfo{person}{Chun-Wei Chiang}, \bibinfo{person}{Zhuoran Lu}, \bibinfo{person}{Zhuoyan Li}, {and} \bibinfo{person}{Ming Yin}.} \bibinfo{year}{2023}\natexlab{}.
\newblock \showarticletitle{Are {Two} {Heads} {Better} {Than} {One} in {AI}-{Assisted} {Decision} {Making}? {Comparing} the {Behavior} and {Performance} of {Groups} and {Individuals} in {Human}-{AI} {Collaborative} {Recidivism} {Risk} {Assessment}}. In \bibinfo{booktitle}{\emph{Proceedings of the 2023 {CHI} {Conference} on {Human} {Factors} in {Computing} {Systems}}} \emph{(\bibinfo{series}{{CHI} '23})}. \bibinfo{publisher}{Association for Computing Machinery}, \bibinfo{address}{New York, NY, USA}, \bibinfo{pages}{1--18}.
\newblock
\showISBNx{978-1-4503-9421-5}
\href{https://doi.org/10.1145/3544548.3581015}{doi:\nolinkurl{10.1145/3544548.3581015}}


\bibitem[Choi and Kim(2026)]%
        {choi_exploring_2026}
\bibfield{author}{\bibinfo{person}{Yoon-Sung Choi} {and} \bibinfo{person}{Min-Cheol Kim}.} \bibinfo{year}{2026}\natexlab{}.
\newblock \showarticletitle{Exploring {Elementary} {Pre}-{Service} {Teachers}’ {AI} {Literacy} in {ChatGPT}-{Assisted} {Science} {Lesson} {Design}}.
\newblock \bibinfo{journal}{\emph{Education Sciences}} \bibinfo{volume}{16}, \bibinfo{number}{8} (\bibinfo{date}{Aug.} \bibinfo{year}{2026}), \bibinfo{pages}{1275}.
\newblock
\shownote{Publisher: Multidisciplinary Digital Publishing Institute}.
\newblock
\showISSN{2227-7102}
\href{https://doi.org/10.3390/educsci16081275}{doi:\nolinkurl{10.3390/educsci16081275}}


\bibitem[Choppin et~al\mbox{.}(2022)]%
        {choppin_role_2022}
\bibfield{author}{\bibinfo{person}{Jeffrey Choppin}, \bibinfo{person}{Amy Roth~McDuffie}, \bibinfo{person}{Corey Drake}, {and} \bibinfo{person}{Jon Davis}.} \bibinfo{year}{2022}\natexlab{}.
\newblock \showarticletitle{The role of instructional materials in the relationship between the official curriculum and the enacted curriculum}.
\newblock \bibinfo{journal}{\emph{Mathematical Thinking and Learning}} \bibinfo{volume}{24}, \bibinfo{number}{2} (\bibinfo{date}{April} \bibinfo{year}{2022}), \bibinfo{pages}{123--148}.
\newblock
\shownote{Publisher: Routledge}.
\newblock
\showISSN{1098-6065}
\href{https://doi.org/10.1080/10986065.2020.1855376}{doi:\nolinkurl{10.1080/10986065.2020.1855376}}


\bibitem[Clarke and Braun(2013)]%
        {clarke_teaching_2013}
\bibfield{author}{\bibinfo{person}{Victoria Clarke} {and} \bibinfo{person}{Virginia Braun}.} \bibinfo{year}{2013}\natexlab{}.
\newblock \showarticletitle{Teaching thematic analysis: {Overcoming} challenges and developing strategies for effective learning}.
\newblock \bibinfo{journal}{\emph{Psychologist}} (\bibinfo{year}{2013}).
\newblock
\urldef\tempurl%
\url{https://www.semanticscholar.org/paper/Teaching-thematic-analysis%3A-Overcoming-challenges-Clarke-Braun/2901ca4762196c5410082d235cda8dd3e1315a08}
\showURL{%
\tempurl}


\bibitem[Cui et~al\mbox{.}(2026)]%
        {cui_codesigning_2026}
\bibfield{author}{\bibinfo{person}{Yuan Cui}, \bibinfo{person}{Annabel~Marie Goldman}, \bibinfo{person}{Jovy Zhou}, \bibinfo{person}{Xiaolin Liu}, \bibinfo{person}{Clarissa~M Shieh}, \bibinfo{person}{Joshua Yao}, \bibinfo{person}{Mia~Lillian Cheng}, \bibinfo{person}{Matthew Kay}, {and} \bibinfo{person}{Fumeng Yang}.} \bibinfo{year}{2026}\natexlab{}.
\newblock \showarticletitle{Codesigning {Ripplet}: an {LLM}-{Assisted} {Assessment} {Authoring} {System} {Grounded} in a {Conceptual} {Model} of {Teachers}’ {Workflows}}. In \bibinfo{booktitle}{\emph{Proceedings of the 2026 {CHI} {Conference} on {Human} {Factors} in {Computing} {Systems}}} \emph{(\bibinfo{series}{{CHI} '26})}. \bibinfo{publisher}{Association for Computing Machinery}, \bibinfo{address}{New York, NY, USA}, \bibinfo{pages}{1--20}.
\newblock
\showISBNx{9798400722783}
\href{https://doi.org/10.1145/3772318.3790418}{doi:\nolinkurl{10.1145/3772318.3790418}}


\bibitem[Czemerinski et~al\mbox{.}(2020)]%
        {czemerinski_whats_2020}
\bibfield{author}{\bibinfo{person}{Hernán Czemerinski}, \bibinfo{person}{Martín Scasso}, {and} \bibinfo{person}{Fernando Schapachnik}.} \bibinfo{year}{2020}\natexlab{}.
\newblock \bibinfo{title}{What's the worth of having a single {CS} teacher program aimed at teachers with heterogeneous profiles?}
\newblock
\shownote{arXiv:2011.04480 [cs.CY]}.
\newblock
\href{https://doi.org/10.48550/arXiv.2011.04480}{doi:\nolinkurl{10.48550/arXiv.2011.04480}}


\bibitem[Fan et~al\mbox{.}(2024)]%
        {fan_lessonplanner_2024}
\bibfield{author}{\bibinfo{person}{Haoxiang Fan}, \bibinfo{person}{Guanzheng Chen}, \bibinfo{person}{Xingbo Wang}, {and} \bibinfo{person}{Zhenhui Peng}.} \bibinfo{year}{2024}\natexlab{}.
\newblock \showarticletitle{{LessonPlanner}: {Assisting} {Novice} {Teachers} to {Prepare} {Pedagogy}-{Driven} {Lesson} {Plans} with {Large} {Language} {Models}}. In \bibinfo{booktitle}{\emph{Proceedings of the 37th {Annual} {ACM} {Symposium} on {User} {Interface} {Software} and {Technology}}} \emph{(\bibinfo{series}{{UIST} '24})}. \bibinfo{publisher}{Association for Computing Machinery}, \bibinfo{address}{New York, NY, USA}, \bibinfo{pages}{1--20}.
\newblock
\showISBNx{9798400706288}
\href{https://doi.org/10.1145/3654777.3676390}{doi:\nolinkurl{10.1145/3654777.3676390}}


\bibitem[Fereday and Muir-Cochrane(2006)]%
        {fereday_demonstrating_2006}
\bibfield{author}{\bibinfo{person}{Jennifer Fereday} {and} \bibinfo{person}{Eimear Muir-Cochrane}.} \bibinfo{year}{2006}\natexlab{}.
\newblock \showarticletitle{Demonstrating {Rigor} {Using} {Thematic} {Analysis}: {A} {Hybrid} {Approach} of {Inductive} and {Deductive} {Coding} and {Theme} {Development}}.
\newblock \bibinfo{journal}{\emph{International Journal of Qualitative Methods: ARCHIVE}} \bibinfo{volume}{5}, \bibinfo{number}{1} (\bibinfo{year}{2006}), \bibinfo{pages}{80--92}.
\newblock
\showISSN{1609-4069}
\urldef\tempurl%
\url{https://journals.library.ualberta.ca/ijqm/index.php/IJQM/article/view/4411}
\showURL{%
\tempurl}


\bibitem[Friend et~al\mbox{.}(2022)]%
        {friend_solve_2022}
\bibfield{author}{\bibinfo{person}{Michelle Friend}, \bibinfo{person}{Monica Mcgill}, {and} \bibinfo{person}{Anni Reinking}.} \bibinfo{year}{2022}\natexlab{}.
\newblock \showarticletitle{Solve {This}! {K}-12 {CS} {Education} {Teachers}’ {Problems} of {Practice}}. In \bibinfo{booktitle}{\emph{Proceedings of the 22nd {Koli} {Calling} {International} {Conference} on {Computing} {Education} {Research}}} \emph{(\bibinfo{series}{Koli {Calling} '22})}. \bibinfo{publisher}{Association for Computing Machinery}, \bibinfo{address}{New York, NY, USA}, \bibinfo{pages}{1--13}.
\newblock
\showISBNx{978-1-4503-9616-5}
\href{https://doi.org/10.1145/3564721.3564738}{doi:\nolinkurl{10.1145/3564721.3564738}}


\bibitem[Gale and Sentance(2026)]%
        {gale_hands-up_2026}
\bibfield{author}{\bibinfo{person}{Laurie Gale} {and} \bibinfo{person}{Sue Sentance}.} \bibinfo{year}{2026}\natexlab{}.
\newblock \showarticletitle{The {Hands}-{Up} {Problem} and {How} to {Deal} {With} {It}: {Secondary} {School} {Teachers}’ {Experiences} of {Debugging} in the {Classroom}}.
\newblock \bibinfo{journal}{\emph{Informatics in Education}} \bibinfo{volume}{25}, \bibinfo{number}{1} (\bibinfo{date}{March} \bibinfo{year}{2026}), \bibinfo{pages}{109--136}.
\newblock
\shownote{Publisher: Vilnius University Institute of Data Science and Digital Technologies}.
\newblock
\showISSN{1648-5831, 2335-8971}
\href{https://doi.org/10.15388/infedu.2504.011}{doi:\nolinkurl{10.15388/infedu.2504.011}}


\bibitem[Gonnermann-Müller et~al\mbox{.}(2026)]%
        {gonnermann-muller_facet_2026}
\bibfield{author}{\bibinfo{person}{Jana Gonnermann-Müller}, \bibinfo{person}{Jennifer Haase}, \bibinfo{person}{Konstantin Fackeldey}, {and} \bibinfo{person}{Sebastian Pokutta}.} \bibinfo{year}{2026}\natexlab{}.
\newblock \bibinfo{title}{{FACET}: {Teacher}-{Centred} {LLM}-{Based} {Multi}-{Agent} {Systems}-{Towards} {Personalized} {Educational} {Worksheets}}.
\newblock
\shownote{arXiv:2508.11401 [cs.HC]}.
\newblock
\href{https://doi.org/10.48550/arXiv.2508.11401}{doi:\nolinkurl{10.48550/arXiv.2508.11401}}


\bibitem[Grave and Pepin(2015)]%
        {grave_teachers_2015}
\bibfield{author}{\bibinfo{person}{Ingvild~Lambert Grave} {and} \bibinfo{person}{Birgit Pepin}.} \bibinfo{year}{2015}\natexlab{}.
\newblock \showarticletitle{Teachers’ use of resources in and for mathematics teaching}.
\newblock \bibinfo{journal}{\emph{NOMAD Nordic Studies in Mathematics Education}} \bibinfo{volume}{20}, \bibinfo{number}{3-4} (\bibinfo{date}{Nov.} \bibinfo{year}{2015}).
\newblock
\showISSN{2002-4061, 1104-2176}
\href{https://doi.org/10.7146/nomad.v20i3-4.148698}{doi:\nolinkurl{10.7146/nomad.v20i3-4.148698}}


\bibitem[Holter and El-Assady(2024)]%
        {holter_deconstructing_2024}
\bibfield{author}{\bibinfo{person}{Steffen Holter} {and} \bibinfo{person}{Mennatallah El-Assady}.} \bibinfo{year}{2024}\natexlab{}.
\newblock \showarticletitle{Deconstructing {Human}-{AI} {Collaboration}: {Agency}, {Interaction}, and {Adaptation}}.
\newblock \bibinfo{journal}{\emph{Computer Graphics Forum}} \bibinfo{volume}{43}, \bibinfo{number}{3} (\bibinfo{year}{2024}), \bibinfo{pages}{e15107}.
\newblock
\showISSN{1467-8659}
\href{https://doi.org/10.1111/cgf.15107}{doi:\nolinkurl{10.1111/cgf.15107}}


\bibitem[Kalenda et~al\mbox{.}(2025)]%
        {kalenda_pre-service_2025}
\bibfield{author}{\bibinfo{person}{Peter~J. Kalenda}, \bibinfo{person}{Logan Rath}, \bibinfo{person}{Marium Abugasea~Heidt}, {and} \bibinfo{person}{Allison Wright}.} \bibinfo{year}{2025}\natexlab{}.
\newblock \showarticletitle{Pre-service {Teacher} {Perceptions} of {ChatGPT} for {Lesson} {Plan} {Generation}}.
\newblock \bibinfo{journal}{\emph{Journal of Educational Technology Systems}} \bibinfo{volume}{53}, \bibinfo{number}{3} (\bibinfo{date}{March} \bibinfo{year}{2025}), \bibinfo{pages}{219--241}.
\newblock
\shownote{Publisher: SAGE Publications Inc}.
\newblock
\showISSN{0047-2395}
\href{https://doi.org/10.1177/00472395241301388}{doi:\nolinkurl{10.1177/00472395241301388}}


\bibitem[Kaputa et~al\mbox{.}(2026)]%
        {kaputa_simstep_2026}
\bibfield{author}{\bibinfo{person}{Zoe Kaputa}, \bibinfo{person}{Anika Rajaram}, \bibinfo{person}{Vryan Feliciano}, \bibinfo{person}{Zhuoyue Lyu}, \bibinfo{person}{Maneesh Agrawala}, {and} \bibinfo{person}{Hariharan Subramonyam}.} \bibinfo{year}{2026}\natexlab{}.
\newblock \showarticletitle{{SimStep}: {Human}-in-the-{Loop} {Authoring} of {Interactive} {Educational} {Simulations} {Through} {Task}-{Level} {Abstractions}}. In \bibinfo{booktitle}{\emph{Proceedings of the 2026 {CHI} {Conference} on {Human} {Factors} in {Computing} {Systems}}} \emph{(\bibinfo{series}{{CHI} '26})}. \bibinfo{publisher}{Association for Computing Machinery}, \bibinfo{address}{New York, NY, USA}, \bibinfo{pages}{1--34}.
\newblock
\showISBNx{9798400722783}
\href{https://doi.org/10.1145/3772318.3791514}{doi:\nolinkurl{10.1145/3772318.3791514}}


\bibitem[Kaufman et~al\mbox{.}(2020)]%
        {kaufman_how_2020}
\bibfield{author}{\bibinfo{person}{Julia~H. Kaufman}, \bibinfo{person}{Sy Doan}, \bibinfo{person}{Andrea Prado~Tuma}, \bibinfo{person}{Ashley Woo}, \bibinfo{person}{Daniella Henry}, {and} \bibinfo{person}{Rebecca~Ann Lawrence}.} \bibinfo{year}{2020}\natexlab{}.
\newblock \bibinfo{booktitle}{\emph{How {Instructional} {Materials} {Are} {Used} and {Supported} in {U}.{S}. {K}–12 {Classrooms}: {Findings} from the 2019 {American} {Instructional} {Resources} {Survey}}}.
\newblock \bibinfo{type}{{T}echnical {R}eport}.
\newblock
\urldef\tempurl%
\url{https://www.rand.org/pubs/research_reports/RRA134-1.html}
\showURL{%
\tempurl}


\bibitem[Kerr and Kim(2025)]%
        {kerr_prompts_2025}
\bibfield{author}{\bibinfo{person}{Robert~C. Kerr} {and} \bibinfo{person}{Hyekyeng Kim}.} \bibinfo{year}{2025}\natexlab{}.
\newblock \showarticletitle{From {Prompts} to {Plans}: {A} {Case} {Study} of {Pre}-{Service} {EFL} {Teachers}’ {Use} of {Generative} {AI} for {Lesson} {Planning}}.
\newblock \bibinfo{journal}{\emph{English Teaching}}  \bibinfo{volume}{80} (\bibinfo{date}{March} \bibinfo{year}{2025}), \bibinfo{pages}{95--118}.
\newblock
\href{https://doi.org/10.15858/engtea.80.1.202503.95}{doi:\nolinkurl{10.15858/engtea.80.1.202503.95}}


\bibitem[Kim et~al\mbox{.}(2021)]%
        {kim_landscape_2021}
\bibfield{author}{\bibinfo{person}{Jiyoung Kim}, \bibinfo{person}{Joshua Childs}, \bibinfo{person}{Anne Leftwich}, \bibinfo{person}{Kendra~Montejos Edwards}, \bibinfo{person}{Carol~L. Fletcher}, {and} \bibinfo{person}{Katie Hendrickson}.} \bibinfo{year}{2021}\natexlab{}.
\newblock \showarticletitle{Landscape of {Computer} {Science} {Teacher} {Qualification} {Pathway}}. In \bibinfo{booktitle}{\emph{Proceedings of the 52nd {ACM} {Technical} {Symposium} on {Computer} {Science} {Education}}} \emph{(\bibinfo{series}{{SIGCSE} '21})}. \bibinfo{publisher}{Association for Computing Machinery}, \bibinfo{address}{New York, NY, USA}, \bibinfo{pages}{1299}.
\newblock
\showISBNx{978-1-4503-8062-1}
\href{https://doi.org/10.1145/3408877.3439635}{doi:\nolinkurl{10.1145/3408877.3439635}}


\bibitem[Lam et~al\mbox{.}(2026)]%
        {lam_just--time_2026}
\bibfield{author}{\bibinfo{person}{Michelle~S. Lam}, \bibinfo{person}{Omar Shaikh}, \bibinfo{person}{Hallie Xu}, \bibinfo{person}{Alice Guo}, \bibinfo{person}{Diyi Yang}, \bibinfo{person}{Jeffrey Heer}, \bibinfo{person}{James~A. Landay}, {and} \bibinfo{person}{Michael~S. Bernstein}.} \bibinfo{year}{2026}\natexlab{}.
\newblock \showarticletitle{Just-{In}-{Time} {Objectives}: {A} {General} {Approach} for {Specialized} {AI} {Interactions}}. In \bibinfo{booktitle}{\emph{Proceedings of the 2026 {CHI} {Conference} on {Human} {Factors} in {Computing} {Systems}}} \emph{(\bibinfo{series}{{CHI} '26})}. \bibinfo{publisher}{Association for Computing Machinery}, \bibinfo{address}{New York, NY, USA}, \bibinfo{pages}{1--26}.
\newblock
\showISBNx{9798400722783}
\href{https://doi.org/10.1145/3772318.3790713}{doi:\nolinkurl{10.1145/3772318.3790713}}


\bibitem[Lawrence et~al\mbox{.}(2024)]%
        {lawrence_how_2024}
\bibfield{author}{\bibinfo{person}{LuEttaMae Lawrence}, \bibinfo{person}{Vanessa Echeverria}, \bibinfo{person}{Kexin Yang}, \bibinfo{person}{Vincent Aleven}, {and} \bibinfo{person}{Nikol Rummel}.} \bibinfo{year}{2024}\natexlab{}.
\newblock \showarticletitle{How teachers conceptualise shared control with an {AI} co-orchestration tool: {A} multiyear teacher-centred design process}.
\newblock \bibinfo{journal}{\emph{British Journal of Educational Technology}} \bibinfo{volume}{55}, \bibinfo{number}{3} (\bibinfo{year}{2024}), \bibinfo{pages}{823--844}.
\newblock
\showISSN{1467-8535}
\href{https://doi.org/10.1111/bjet.13372}{doi:\nolinkurl{10.1111/bjet.13372}}


\bibitem[Lehtinen et~al\mbox{.}(2026)]%
        {lehtinen_generative_2026}
\bibfield{author}{\bibinfo{person}{Auli Lehtinen}, \bibinfo{person}{Faisal Channa}, \bibinfo{person}{Piia Näykki}, \bibinfo{person}{Emilia Ahlström}, {and} \bibinfo{person}{Mikko Hiljanen}.} \bibinfo{year}{2026}\natexlab{}.
\newblock \showarticletitle{Generative {AI} for {Collaborative} {Learning}: {Fostering} {Critical} {Thinking} in {Teacher} {Education}}.
\newblock \bibinfo{journal}{\emph{Journal of Computer Assisted Learning}} \bibinfo{volume}{42}, \bibinfo{number}{3} (\bibinfo{year}{2026}), \bibinfo{pages}{e70256}.
\newblock
\showISSN{1365-2729}
\href{https://doi.org/10.1002/jcal.70256}{doi:\nolinkurl{10.1002/jcal.70256}}


\bibitem[Li et~al\mbox{.}(2023)]%
        {li_modeling_2023}
\bibfield{author}{\bibinfo{person}{Zhuoyan Li}, \bibinfo{person}{Zhuoran Lu}, {and} \bibinfo{person}{Ming Yin}.} \bibinfo{year}{2023}\natexlab{}.
\newblock \showarticletitle{Modeling human trust and reliance in {AI}-assisted decision making: a markovian approach}. In \bibinfo{booktitle}{\emph{Proceedings of the {Thirty}-{Seventh} {AAAI} {Conference} on {Artificial} {Intelligence} and {Thirty}-{Fifth} {Conference} on {Innovative} {Applications} of {Artificial} {Intelligence} and {Thirteenth} {Symposium} on {Educational} {Advances} in {Artificial} {Intelligence}}} \emph{(\bibinfo{series}{{AAAI}'23/{IAAI}'23/{EAAI}'23}, Vol.~\bibinfo{volume}{37})}. \bibinfo{publisher}{AAAI Press}, \bibinfo{pages}{6056--6064}.
\newblock
\showISBNx{978-1-57735-880-0}
\href{https://doi.org/10.1609/aaai.v37i5.25748}{doi:\nolinkurl{10.1609/aaai.v37i5.25748}}


\bibitem[Ma et~al\mbox{.}(2025)]%
        {ma_towards_2025}
\bibfield{author}{\bibinfo{person}{Shuai Ma}, \bibinfo{person}{Qiaoyi Chen}, \bibinfo{person}{Xinru Wang}, \bibinfo{person}{Chengbo Zheng}, \bibinfo{person}{Zhenhui Peng}, \bibinfo{person}{Ming Yin}, {and} \bibinfo{person}{Xiaojuan Ma}.} \bibinfo{year}{2025}\natexlab{}.
\newblock \showarticletitle{Towards {Human}-{AI} {Deliberation}: {Design} and {Evaluation} of {LLM}-{Empowered} {Deliberative} {AI} for {AI}-{Assisted} {Decision}-{Making}}. In \bibinfo{booktitle}{\emph{Proceedings of the 2025 {CHI} {Conference} on {Human} {Factors} in {Computing} {Systems}}} \emph{(\bibinfo{series}{{CHI} '25})}. \bibinfo{publisher}{Association for Computing Machinery}, \bibinfo{address}{New York, NY, USA}, \bibinfo{pages}{1--23}.
\newblock
\showISBNx{9798400713941}
\href{https://doi.org/10.1145/3706598.3713423}{doi:\nolinkurl{10.1145/3706598.3713423}}


\bibitem[Ni et~al\mbox{.}(2023)]%
        {ni_computer_2023}
\bibfield{author}{\bibinfo{person}{Lijun Ni}, \bibinfo{person}{Gillian Bausch}, {and} \bibinfo{person}{Rebecca Benjamin}.} \bibinfo{year}{2023}\natexlab{}.
\newblock \showarticletitle{Computer science teacher professional development and professional learning communities: a review of the research literature}.
\newblock \bibinfo{journal}{\emph{Computer Science Education}} \bibinfo{volume}{33}, \bibinfo{number}{1} (\bibinfo{date}{Jan.} \bibinfo{year}{2023}), \bibinfo{pages}{29--60}.
\newblock
\shownote{Publisher: Routledge}.
\newblock
\showISSN{0899-3408}
\href{https://doi.org/10.1080/08993408.2021.1993666}{doi:\nolinkurl{10.1080/08993408.2021.1993666}}


\bibitem[Pareek et~al\mbox{.}(2026)]%
        {pareek_sensemaking_2026}
\bibfield{author}{\bibinfo{person}{Saumya Pareek}, \bibinfo{person}{Jarod Govers}, \bibinfo{person}{Naja~Kathrine Kollerup}, \bibinfo{person}{Emily Wong}, \bibinfo{person}{Eduardo Velloso}, {and} \bibinfo{person}{Jorge Goncalves}.} \bibinfo{year}{2026}\natexlab{}.
\newblock \showarticletitle{Sensemaking in {Multi}-{Agent} {LLM} {Interfaces}: {How} {Users} {Interpret} {Transparency} and {Trustworthiness} {Cues}}. In \bibinfo{booktitle}{\emph{Proceedings of the 2026 {CHI} {Conference} on {Human} {Factors} in {Computing} {Systems}}} \emph{(\bibinfo{series}{{CHI} '26})}. \bibinfo{publisher}{Association for Computing Machinery}, \bibinfo{address}{New York, NY, USA}, \bibinfo{pages}{1--20}.
\newblock
\showISBNx{9798400722783}
\href{https://doi.org/10.1145/3772318.3791157}{doi:\nolinkurl{10.1145/3772318.3791157}}


\bibitem[Polikoff(2018)]%
        {polikoff_challenges_2018}
\bibfield{author}{\bibinfo{person}{Morgan Polikoff}.} \bibinfo{year}{2018}\natexlab{}.
\newblock \bibinfo{title}{The challenges of curriculum materials as a reform lever}.
\newblock
\urldef\tempurl%
\url{https://www.brookings.edu/articles/the-challenges-of-curriculum-materials-as-a-reform-lever/}
\showURL{%
\tempurl}


\bibitem[Polikoff(2019)]%
        {polikoff_supplemental_2019}
\bibfield{author}{\bibinfo{person}{Morgan Polikoff}.} \bibinfo{year}{2019}\natexlab{}.
\newblock \bibinfo{booktitle}{\emph{The {Supplemental} {Curriculum} {Bazaar}: {Is} {What}'s {Online} {Any} {Good}?}}
\newblock \bibinfo{type}{{T}echnical {R}eport}. \bibinfo{institution}{Thomas B}.
\newblock
\shownote{Publication Title: Thomas B. Fordham Institute ERIC Number: ED601253}.
\newblock
\urldef\tempurl%
\url{https://eric.ed.gov/?id=ED601253}
\showURL{%
\tempurl}


\bibitem[Polikoff and Silver(2021)]%
        {polikoff_identifying_2021}
\bibfield{author}{\bibinfo{person}{Morgan~S. Polikoff} {and} \bibinfo{person}{Daniel Silver}.} \bibinfo{year}{2021}\natexlab{}.
\newblock \showarticletitle{Identifying and {Distinguishing} {Among} {Teachers}’ {Supplementary} {Curriculum} {Use} {Patterns} {Using} the {Lasso}}.
\newblock \bibinfo{journal}{\emph{Frontiers in Education}}  \bibinfo{volume}{6} (\bibinfo{date}{Aug.} \bibinfo{year}{2021}).
\newblock
\shownote{Publisher: Frontiers}.
\newblock
\showISSN{2504-284X}
\href{https://doi.org/10.3389/feduc.2021.722554}{doi:\nolinkurl{10.3389/feduc.2021.722554}}


\bibitem[Reichert et~al\mbox{.}(2027)]%
        {reichert_framework_2027}
\bibfield{author}{\bibinfo{person}{Heidi Reichert}, \bibinfo{person}{Tahreem Yasir}, \bibinfo{person}{Malvika Satyavolu}, {and} \bibinfo{person}{Tiffany Barnes}.} \bibinfo{year}{2027}\natexlab{}.
\newblock \showarticletitle{A {Framework} for {LLM} {Integration} in {Secondary} {Education}: {Insights} from {Computing} {Teachers}}. In \bibinfo{booktitle}{\emph{Artificial {Intelligence} in {Education}}}, \bibfield{editor}{\bibinfo{person}{Emmanuel~G. Blanchard}, \bibinfo{person}{Guanliang Chen}, \bibinfo{person}{Min Chi}, {and} \bibinfo{person}{Seiji Isotani}} (Eds.). \bibinfo{publisher}{Springer Nature Switzerland}, \bibinfo{address}{Cham}, \bibinfo{pages}{107--121}.
\newblock
\showISBNx{978-3-032-29763-1}
\href{https://doi.org/10.1007/978-3-032-29763-1_8}{doi:\nolinkurl{10.1007/978-3-032-29763-1_8}}


\bibitem[Rhodes(2018)]%
        {rhodes_design_2018}
\bibfield{author}{\bibinfo{person}{Holly~G. Rhodes}.} \bibinfo{year}{2018}\natexlab{}.
\newblock \bibinfo{booktitle}{\emph{Design, {Selection}, and {Implementation} of {Instructional} {Materials} for the {Next} {Generation} {Science} {Standards}}}.
\newblock
\urldef\tempurl%
\url{https://www.nationalacademies.org/read/25001/chapter/5}
\showURL{%
\tempurl}


\bibitem[Sharma et~al\mbox{.}(2024)]%
        {sharma_investigating_2024}
\bibfield{author}{\bibinfo{person}{Ashish Sharma}, \bibinfo{person}{Sudha Rao}, \bibinfo{person}{Chris Brockett}, \bibinfo{person}{Akanksha Malhotra}, \bibinfo{person}{Nebojsa Jojic}, {and} \bibinfo{person}{Bill Dolan}.} \bibinfo{year}{2024}\natexlab{}.
\newblock \showarticletitle{Investigating {Agency} of {LLMs} in {Human}-{AI} {Collaboration} {Tasks}}. In \bibinfo{booktitle}{\emph{Proceedings of the 18th {Conference} of the {European} {Chapter} of the {Association} for {Computational} {Linguistics} ({Volume} 1: {Long} {Papers})}}, \bibfield{editor}{\bibinfo{person}{Yvette Graham} {and} \bibinfo{person}{Matthew Purver}} (Eds.). \bibinfo{publisher}{Association for Computational Linguistics}, \bibinfo{address}{St. Julian's, Malta}, \bibinfo{pages}{1968--1987}.
\newblock
\href{https://doi.org/10.18653/v1/2024.eacl-long.119}{doi:\nolinkurl{10.18653/v1/2024.eacl-long.119}}


\bibitem[Sherin and Drake(2009)]%
        {sherin_curriculum_2009}
\bibfield{author}{\bibinfo{person}{Miriam~Gamoran Sherin} {and} \bibinfo{person}{Corey Drake}.} \bibinfo{year}{2009}\natexlab{}.
\newblock \showarticletitle{Curriculum strategy framework: investigating patterns in teachers’ use of a reform‐based elementary mathematics curriculum}.
\newblock \bibinfo{journal}{\emph{Journal of Curriculum Studies}} \bibinfo{volume}{41}, \bibinfo{number}{4} (\bibinfo{date}{Aug.} \bibinfo{year}{2009}), \bibinfo{pages}{467--500}.
\newblock
\shownote{Publisher: Routledge}.
\newblock
\showISSN{0022-0272}
\href{https://doi.org/10.1080/00220270802696115}{doi:\nolinkurl{10.1080/00220270802696115}}


\bibitem[Subramonyam et~al\mbox{.}(2024)]%
        {subramonyam_bridging_2024}
\bibfield{author}{\bibinfo{person}{Hari Subramonyam}, \bibinfo{person}{Roy Pea}, \bibinfo{person}{Christopher Pondoc}, \bibinfo{person}{Maneesh Agrawala}, {and} \bibinfo{person}{Colleen Seifert}.} \bibinfo{year}{2024}\natexlab{}.
\newblock \showarticletitle{Bridging the {Gulf} of {Envisioning}: {Cognitive} {Challenges} in {Prompt} {Based} {Interactions} with {LLMs}}. In \bibinfo{booktitle}{\emph{Proceedings of the 2024 {CHI} {Conference} on {Human} {Factors} in {Computing} {Systems}}} \emph{(\bibinfo{series}{{CHI} '24})}. \bibinfo{publisher}{Association for Computing Machinery}, \bibinfo{address}{New York, NY, USA}, \bibinfo{pages}{1--19}.
\newblock
\showISBNx{9798400703300}
\href{https://doi.org/10.1145/3613904.3642754}{doi:\nolinkurl{10.1145/3613904.3642754}}


\bibitem[Varuvel~Dennison et~al\mbox{.}(2026)]%
        {varuvel_dennison_shiksha_2026}
\bibfield{author}{\bibinfo{person}{Deepak Varuvel~Dennison}, \bibinfo{person}{Bakhtawar Ahtisham}, \bibinfo{person}{Kavyansh Chourasia}, \bibinfo{person}{Nirmit Arora}, \bibinfo{person}{Rahul Singh}, \bibinfo{person}{René~F. Kizilcec}, \bibinfo{person}{Akshay Nambi}, \bibinfo{person}{Tanuja Ganu}, {and} \bibinfo{person}{Aditya Vashistha}.} \bibinfo{year}{2026}\natexlab{}.
\newblock \showarticletitle{Shiksha {Copilot}: {Teacher}-{AI} {Collaboration} for {Curating} and {Customizing} {Lesson} {Plans} in {Low}-{Resource} {Schools}}.
\newblock \bibinfo{journal}{\emph{Proc. ACM Hum.-Comput. Interact.}} \bibinfo{volume}{10}, \bibinfo{number}{2} (\bibinfo{year}{2026}), \bibinfo{pages}{CSCW038:1--CSCW038:47}.
\newblock
\href{https://doi.org/10.1145/3788074}{doi:\nolinkurl{10.1145/3788074}}


\bibitem[Vinnervik(2023)]%
        {vinnervik_-depth_2023}
\bibfield{author}{\bibinfo{person}{Peter Vinnervik}.} \bibinfo{year}{2023}\natexlab{}.
\newblock \showarticletitle{An in-depth analysis of programming in the {Swedish} school curriculum—rationale, knowledge content and teacher guidance}.
\newblock \bibinfo{journal}{\emph{Journal of Computers in Education}} \bibinfo{volume}{10}, \bibinfo{number}{2} (\bibinfo{date}{June} \bibinfo{year}{2023}), \bibinfo{pages}{237--271}.
\newblock
\showISSN{2197-9995}
\href{https://doi.org/10.1007/s40692-022-00230-2}{doi:\nolinkurl{10.1007/s40692-022-00230-2}}


\bibitem[Wang et~al\mbox{.}(2021)]%
        {wang_teachers_2021}
\bibfield{author}{\bibinfo{person}{Elaine~Lin Wang}, \bibinfo{person}{Andrea Prado~Tuma}, \bibinfo{person}{Sy Doan}, \bibinfo{person}{Daniella Henry}, \bibinfo{person}{Rebecca~Ann Lawrence}, \bibinfo{person}{Ashley Woo}, {and} \bibinfo{person}{Julia~H. Kaufman}.} \bibinfo{year}{2021}\natexlab{}.
\newblock \bibinfo{booktitle}{\emph{Teachers' {Perceptions} of {What} {Makes} {Instructional} {Materials} {Engaging}, {Appropriately} {Challenging}, and {Usable}: {A} {Survey} and {Interview} {Study}}}.
\newblock \bibinfo{type}{{T}echnical {R}eport}.
\newblock
\urldef\tempurl%
\url{https://www.rand.org/pubs/research_reports/RRA134-2.html}
\showURL{%
\tempurl}


\bibitem[Wang et~al\mbox{.}(2025)]%
        {wang_enabling_2025}
\bibfield{author}{\bibinfo{person}{Jiayi Wang}, \bibinfo{person}{Ruiwei Xiao}, \bibinfo{person}{Xinying Hou}, {and} \bibinfo{person}{John Stamper}.} \bibinfo{year}{2025}\natexlab{}.
\newblock \bibinfo{title}{Enabling {Multi}-{Agent} {Systems} as {Learning} {Designers}: {Applying} {Learning} {Sciences} to {AI} {Instructional} {Design}}.
\newblock
\shownote{arXiv:2508.16659 [cs.CY]}.
\newblock
\href{https://doi.org/10.48550/arXiv.2508.16659}{doi:\nolinkurl{10.48550/arXiv.2508.16659}}


\bibitem[Wilson et~al\mbox{.}(2010)]%
        {wilson_running_2010}
\bibfield{author}{\bibinfo{person}{Cameron Wilson}, \bibinfo{person}{Leigh~Ann Sudol}, \bibinfo{person}{Chris Stephenson}, {and} \bibinfo{person}{Mark Stehlik}.} \bibinfo{year}{2010}\natexlab{}.
\newblock \bibinfo{booktitle}{\emph{Running on {Empty}: the {Failure} to {Teach} {K}--12 {Computer} {Science} in the {Digital} {Age}}}.
\newblock \bibinfo{publisher}{Association for Computing Machinery}, \bibinfo{address}{New York, NY, USA}.
\newblock
\showISBNx{978-1-4503-8867-2}
\urldef\tempurl%
\url{https://dl.acm.org/doi/10.1145/3414583}
\showURL{%
\tempurl}


\bibitem[Wu et~al\mbox{.}(2022)]%
        {wu_ai_2022}
\bibfield{author}{\bibinfo{person}{Tongshuang Wu}, \bibinfo{person}{Michael Terry}, {and} \bibinfo{person}{Carrie~Jun Cai}.} \bibinfo{year}{2022}\natexlab{}.
\newblock \showarticletitle{{AI} {Chains}: {Transparent} and {Controllable} {Human}-{AI} {Interaction} by {Chaining} {Large} {Language} {Model} {Prompts}}. In \bibinfo{booktitle}{\emph{Proceedings of the 2022 {CHI} {Conference} on {Human} {Factors} in {Computing} {Systems}}} \emph{(\bibinfo{series}{{CHI} '22})}. \bibinfo{publisher}{Association for Computing Machinery}, \bibinfo{address}{New York, NY, USA}, \bibinfo{pages}{1--22}.
\newblock
\showISBNx{978-1-4503-9157-3}
\href{https://doi.org/10.1145/3491102.3517582}{doi:\nolinkurl{10.1145/3491102.3517582}}


\bibitem[Yizhou~Tian et~al\mbox{.}(2026)]%
        {yizhou_tian_understanding_2026}
\bibfield{author}{\bibinfo{person}{Harry Yizhou~Tian}, \bibinfo{person}{Hasan Amin}, {and} \bibinfo{person}{Ming Yin}.} \bibinfo{year}{2026}\natexlab{}.
\newblock \showarticletitle{Understanding the {Effects} of {AI}-{Assisted} {Critical} {Thinking} on {Human}-{AI} {Decision} {Making}}. In \bibinfo{booktitle}{\emph{Proceedings of the 2026 {CHI} {Conference} on {Human} {Factors} in {Computing} {Systems}}} \emph{(\bibinfo{series}{{CHI} '26})}. \bibinfo{publisher}{Association for Computing Machinery}, \bibinfo{address}{New York, NY, USA}, \bibinfo{pages}{1--30}.
\newblock
\showISBNx{9798400722783}
\href{https://doi.org/10.1145/3772318.3790785}{doi:\nolinkurl{10.1145/3772318.3790785}}


\bibitem[Yu et~al\mbox{.}(2026)]%
        {yu_improving_2026}
\bibfield{author}{\bibinfo{person}{Jiaen Yu}, \bibinfo{person}{Ylesia Wu}, \bibinfo{person}{Gabriel Cha}, \bibinfo{person}{Ayush Shah}, {and} \bibinfo{person}{Samuel Lau}.} \bibinfo{year}{2026}\natexlab{}.
\newblock \showarticletitle{Improving {LLM}-{Generated} {Educational} {Content}: {A} {Case} {Study} on {Prototyping}, {Prompt} {Engineering}, and {Evaluating} a {Tool} for {Generating} {Programming} {Problems} for {Data} {Science}}. In \bibinfo{booktitle}{\emph{Proceedings of the 57th {ACM} {Technical} {Symposium} on {Computer} {Science} {Education} {V}.1}} \emph{(\bibinfo{series}{{SIGCSE} {TS} 2026}, Vol.~\bibinfo{volume}{1})}. \bibinfo{publisher}{Association for Computing Machinery}, \bibinfo{address}{New York, NY, USA}, \bibinfo{pages}{1186--1192}.
\newblock
\showISBNx{9798400722561}
\href{https://doi.org/10.1145/3770762.3772619}{doi:\nolinkurl{10.1145/3770762.3772619}}


\bibitem[Zhang et~al\mbox{.}(2025)]%
        {zhang_introducing_2025}
\bibfield{author}{\bibinfo{person}{Shan Zhang}, \bibinfo{person}{Nicole~M. Hutchins}, \bibinfo{person}{Joanne~R. Barrett}, \bibinfo{person}{Anthony~F. Botelho}, {and} \bibinfo{person}{Maya Israel}.} \bibinfo{year}{2025}\natexlab{}.
\newblock \showarticletitle{Introducing {K}-12 {Teachers} to {Computer} {Science} {Education} through an {Online} {Micro}-credential: {An} {Experience} {Report}}. In \bibinfo{booktitle}{\emph{Proceedings of the 56th {ACM} {Technical} {Symposium} on {Computer} {Science} {Education} {V}. 1}} \emph{(\bibinfo{series}{{SIGCSETS} 2025})}. \bibinfo{publisher}{Association for Computing Machinery}, \bibinfo{address}{New York, NY, USA}, \bibinfo{pages}{1316--1322}.
\newblock
\showISBNx{9798400705311}
\href{https://doi.org/10.1145/3641554.3701855}{doi:\nolinkurl{10.1145/3641554.3701855}}


\bibitem[Zhang et~al\mbox{.}(2026)]%
        {zhang_adapt_2026}
\bibfield{author}{\bibinfo{person}{Yanjie Zhang}, \bibinfo{person}{Jiajun Zhu}, \bibinfo{person}{Minyu Wu}, \bibinfo{person}{Huamin Qu}, {and} \bibinfo{person}{Sicheng Song}.} \bibinfo{year}{2026}\natexlab{}.
\newblock \bibinfo{title}{{AdaPT}: {Adaptive} {Lesson} {Plan} {Transformer} for {Cross}-{Regional} and {Differentiated} {Instruction}}.
\newblock
\shownote{arXiv:2606.17633 [cs.HC]}.
\newblock
\href{https://doi.org/10.48550/arXiv.2606.17633}{doi:\nolinkurl{10.48550/arXiv.2606.17633}}


\end{thebibliography}
